\documentclass{article}

\usepackage{graphicx} 
\usepackage{lmodern}

\usepackage{tabularx}
\usepackage{array}
\usepackage{amsmath} 

\usepackage[hidelinks]{hyperref}
\usepackage{authblk}
\usepackage{orcidlink}
\usepackage[a4paper, margin=1in]{geometry}

\begin{document}

\title{Impact of Exchange-Correlation Functionals on Predictions of Phonon Hydrodynamics: A Study of Fluorides, Chlorides, and Hydrides}

\author{Jamal Abou Haibeh$^1$~\orcidlink{0009-0003-7167-4731} and Samuel Huberman$^{1,2,*}$~\orcidlink{0000-0003-0865-8096}\thanks{Corresponding author: \href{samuel.huberman@mcgill.ca}{samuel.huberman@mcgill.ca}}}

\affil{$^1$Department of Chemical Engineering, McGill University, Montreal, Quebec H3A 0C5, Canada}

\affil{$^2$Department of Physics, McGill University, Montreal, Quebec H3A 0C5, Canada}

\date{}

\maketitle

\begin{abstract}
We employ density functional theory calculations to examine the effect of various exchange-correlation (XC) functionals, including the Perdew-Burke-Ernzerhof generalized gradient approximation (PBE), the modified Perdew-Burke-Ernzerhof generalized gradient approximation (PBEsol), and the local density approximation (LDA), on the electrical, mechanical, and thermal properties of sodium fluoride (NaF), lithium fluoride (LiF), potassium fluoride (KF), sodium chloride (NaCl), potassium chloride (KCl), lithium hydride (LiH), sodium hydride (NaH), and potassium hydride (KH). The lattice thermal conductivity is computed based on an iterative solution of the Boltzmann transport equation (BTE). Based on Guyer's criterion and direct solutions to the linearized BTE, we determine the ballistic, phonon hydrodynamics, and diffusive regimes as a function of length scale and temperature. In addition to confirming previous predictions of phonon hydrodynamics in NaF and LiF, we report novel predictions of phonon hydrodynamics in NaH,  LiH, KH, KF, NaCl, and KCl. The impact of isotopes on the calculated lattice thermal conductivity and phonon hydrodynamics windows is also reported. The impact of Meta-GGA and hybrid functionals is also discussed. We find that the selection of a functional impacts the prediction of thermal conductivity and the window for observation of phonon hydrodynamics.

\end{abstract}

\section{Introduction}

Based on the original insights of Hohenberg and Kohn, who first proved that the ground-state electron density uniquely determines the energy of a many-electron system \cite{kohn1965self}, density functional theory (DFT) has emerged as a standard tool in computational materials science. DFT's success can be attributed to its ability to balance accuracy and computational cost \cite{van2014density}. At the heart of this balance lies the exchange-correlation (XC) functional, which approximates many-body electron interactions. This functional operates on a foundation that facilitates a ladder of approximations. The choice of XC approximation is critical, as it largely determines the trade-off between accuracy and computational expense in DFT simulations \cite{dick2020machine}.

The bottom rung in the ladder of functional approximations is the local density approximation (LDA), which approximates the exchange correlation energy based on a uniform electron gas model \cite{perdew1996generalized}. Advancements led to the formulation of generalized gradient approximations (GGA), which extend the LDA approach by incorporating gradient corrections to the electron density, enhancing the accuracy of DFT calculations. Several GGA functionals have been developed, as there are multiple ways to characterize the gradient of the electron density  \cite{perdew1996generalized, chen2006b3lyp, perdew2008restoring}. The two most common GGA functionals are the Perdew-Burke-Ernzerhof generalized gradient approximation (PBE) \cite{perdew1996generalized} and the modified Perdew-Burke-Ernzerhof generalized gradient approximation (PBEsol) \cite{perdew2008restoring}. PBE includes the gradient of the electron density in the exchange correlation function and improves upon LDA by capturing some non-local effects, but there remain limitations, particularly for systems with strongly localized electrons (e.g., transition metal complexes \cite{zhang2018performance}). The third most common functional is meta-generalized gradient approximation (meta-GGA), which augments the functional dependence by including higher-order terms like the second derivative of the electron density \cite{marques2003primer}. In general, meta-GGAs (e.g., SCAN) offer improvements over GGAs in terms of accuracy for a variety of properties, including thermochemistry, and provide reliable predictions of structures and energies for a wide range of molecular and material bonding types \cite{sun2016accurate}. Another noteworthy functional is the HSE (Heyd-Scuseria-Ernzerhof) hybrid functional which combines the traditional DFT exchange-correlation functional with a fraction of Hartree-Fock (HF) exchange \cite{heyd2003hybrid} via a mixing parameter that determines the contribution of HF exchange to the total exchange-correlation energy. However, this functional requires greater computational resources than the simple PBE functional.

Given these approximations, deviations between experimental and theoretical predictions are expected. It is well known that the LDA under-predicts the bandgap in silicon by about 50 \% \cite{persson2006improved,ishii2010all,zhao2019prediction}. However, even with this significant discrepancy, accurate DFT-based predictions of the thermal conductivity of silicon have been obtained \cite{lindsay2013first, parks2020uncertainty, mortazavi2021accelerating}. For instance, Jain and McGaughey showed that the choice of pseudopotential and XC functional in thermal conductivity predictions can impact accuracy by up to 17\% \cite{jain2015effect}. 

Going beyond thermal conductivity, recent work has built upon these first principles calculations to investigate phonon hydrodynamics, a regime of thermal transport that lies between the diffusive and ballistic regimes \cite{ghosh2022phonon}. One of the signatures of phonon hydrodynamics is the``second sound'', where the temperature propagates as a wave. Early observations of second sound were obtained for solid He$^4$ ~\cite{ackerman1966second}, solid He$^3$ ~\cite{ackerman1969second} and NaF ~\cite{jackson1970second}. Subsequent observations of second sound are scarce and limited to bismuth~\cite{narayanamurti1972observation} and graphite ~\cite{huberman2019observation, ding2022observation}. Beyond these experimental observations of phonon hydrodynamics, there is a paucity of other experimental work. Rogers ~\cite{rogers1971transport} identified a narrow intermediate-temperature window in \({}^{7}\mathrm{LiF}\), where normal processes shape the propagated heat pulse, but the experimental evidence for phonon hydrodynamics remains inconclusive. Jelinek \textit{et\,al} ~\cite{jelinek1987thermal} measured the thermal conductivity of ${}^{6}\mathrm{LiH}$ with 0.7\% porosity from 6 to 399\,K and observed a pronounced peak at approximately 50--70\,K. Guyer and Krumhansl ~\cite{guyer1966thermal} argued that isotope scattering in the then-available LiF crystals makes Poiseuille-type hydrodynamic flow unlikely, but second sound could still be observed at around 20 K in 2 mm sample radius, high-purity samples. Slack ~\cite{slack1957thermal} demonstrated that thermal conductivity in pure KCl at low temperatures is limited by boundary and isotope scattering, but no statement about the possibility of phonon hydrodynamics is made. A similar experimental and theoretical analysis was done for NaCl ~\cite{caldwell1967experimental}. To date, DFT has not been used to study phonon hydrodynamics in these materials. The objective of this work is to examine the impact of the choice of XC functionals (PBE, PBEsol, LDA) on the prediction of phonon hydrodynamics in fluorides, chlorides, and hydrides.

\section{Theory}

Based on approximate solutions to the phonon Boltzmann Transport Equation (BTE), a condition for the existence of phonon hydrodynamics was proposed by Guyer and Krumhansl, henceforth referred to as Guyer’s criterion \cite{guyer1966thermal}. This condition states that phonon hydrodynamics can be observed if the average boundary scattering rate $\langle\tau_B^{-1}\rangle$ is greater than the average Umklapp scattering rate $\langle\tau_U^{-1}\rangle$ but less than the average Normal scattering rate $\langle\tau_N^{-1}\rangle$ at a given temperature and for a characteristic length \cite{guyer1966thermal}: 

\begin{equation}
\langle\tau_{N}^{-1}\rangle \;>\;\langle\tau_{B}^{-1}\rangle \;>\;\langle\tau_{U}^{-1}\rangle,
\label{eq:guyers-criterion}
\end{equation}

Here, $N$ denotes Normal (momentum-conserving) phonon–phonon scattering, $U$ denotes Umklapp (momentum-relaxing) phonon–phonon scattering, and $B$ denotes boundary scattering. In other words, phonon hydrodynamics can be observed if Normal scattering dominates over boundary scattering and Umklapp scattering.  The average resistive scattering rate $\langle\tau_{R}^{-1}\rangle$ (sum of Umklapp and isotope scattering rates) can also be used instead of the average Umklapp scattering rate:

\begin{equation}
\langle\tau_{N}^{-1}\rangle \;>\;\langle\tau_{B}^{-1}\rangle \;>\;\langle\tau_{R}^{-1}\rangle,
\label{eq:resistive-criterion}
\end{equation}

where $\langle\cdot\rangle$ denotes the average over the Brillouin zone.  The average scattering rate $\langle\tau^{-1}\rangle$ is defined as:

\begin{equation}
\bigl\langle \tau^{-1}\bigr\rangle
= \frac{\sum_{i=1}^N \tau_i^{-1}\,C_i}{\sum_{i=1}^N C_i}
= \frac{\sum_{i=1}^N \tau_i^{-1}\,C_i}{C},
\label{eq:avg-scattering-rate}
\end{equation}

where \(C_i\) and \(C\) are the heat capacity for a given phonon mode $i$ and the total heat capacity of all phonon modes and $N$ is the number of q-points in the discretized Brillouin zone, respectively.  Other expressions for average scattering rates may be selected, but here we follow previous studies \cite{cepellotti2015phonon, ghosh2020phonon, ghosh2022phonon}. Extending our earlier study on NaF, LiF, LiH, and NaH \cite{abou2022phonon, huberman2023first}, we expand the analysis to a broader set of eight rock-salt compounds by including four additional materials and providing a more detailed, systematic comparison of XC functional effects on phonon transport and hydrodynamic windows. Details of the DFT and BTE calculations can be found in Appendix A of the supplementary material.

\section{Results}

\subsection{General Properties}

\begin{figure}[ht]
  \centering
  \includegraphics[width=0.5\linewidth]{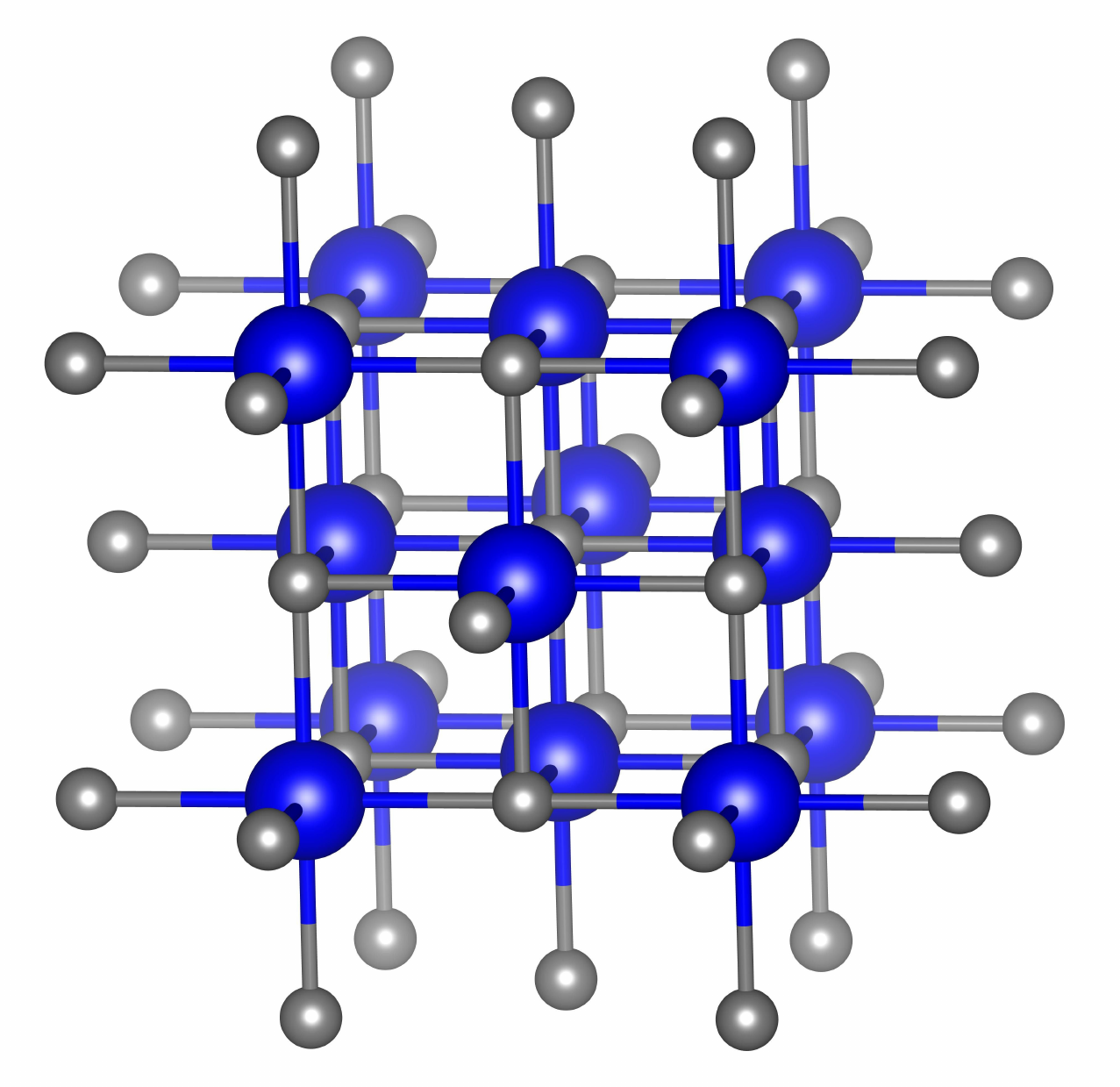}
  \caption{Unit cell of NaCl cubic crystal. Blue spheres denote Na atoms; gray spheres denote Cl atoms.}
  \label{fig:crystal}
\end{figure}

All eight compounds studied in this work (NaF, LiF, KF, NaCl, KCl, LiH, NaH, and KH) crystallize in the rock-salt structure (face-centered cubic, space group $Fm\overline{3}m$)  \cite{hoffmann2020introduction}. Each cation and anion is octahedrally coordinated by six neighbors. NaF, LiF, KF, KCl, LiH, NaH, and KH all adopt the NaCl-type lattice with similar octahedral frameworks, as shown in figure ~\ref{fig:crystal}. A comprehensive analysis, given in table~\ref{tab:general-materials}, of the structural, dielectric, vibrational, and electronic properties highlights the relative performance of the PBE, PBEsol, and LDA functionals. The calculated results are systematically compared against available experimental and theoretical data. For fluorides and chlorides, PBE slightly overestimates and LDA underestimates the lattice constant, while PBEsol is the closest to experiments, often deviating by less than 1\% from experimental values. As GGA includes density-gradient corrections that reduce LDA’s systematic overbinding, it typically yields more accurate equilibrium volumes and lattice constants for real solids. For instance, PBEsol lattice constants for KF (5.31 \AA) and KCl (6.24 \AA) are in excellent agreement with their experimental measurement results of 5.31~\AA{} and 6.25~\AA{} \cite{pies2006crystal}. Conversely, for the hydrides, the PBE functional tends to yield better results, as seen with LiH, where the PBE value (4.01 \AA{}) is the closest to the experimental value of 4.08 \AA{} \cite{pretzel1960properties}. The lattice constant increases monotonically from Li to Na to K for each anion, corresponding to the increasing ionic radii.

The dielectric-constant tensor and Born effective charges are calculated based on the density functional perturbation theory (DFPT) to capture the longitudinal and transverse optical phonon (LO-TO) splitting at the gamma point \cite{liang2018lattice}. For fluorides, chlorides and hydrides (NaF, LiF, LiH, etc.), the underlying Bravais lattice is cubic; hence, there is only one independent component in the Born effective charge tensor. In all cases, Born effective charges are nearly ionic ($\lvert Z^{\ast}\rvert \approx 1$), exemplified by NaCl ($1.11,\,1.09,\,1.07$ vs.\ $1.099$ \cite{shukla2000ab}) and LiH ($1.02$--$1.03$ vs.\ $1.11$ \cite{blat1991calculations}). For non-magnetic, transparent solids, a larger electronic polarizability leads to a larger refractive index ($n$) according to the Lorentz--Lorenz relation: $(n^{2}-1)/(n^{2}+2)=N\alpha/(3\varepsilon_{0})$, and in the optical limit with $\mu_r\approx 1$ the high-frequency dielectric constant satisfies $\varepsilon_{\infty}\approx n^{2}$ \cite{griffiths2023introduction}. Across anions at fixed cation, the electronic polarizability increases in the sequence $\mathrm{F^-} < \mathrm{Cl^-} < \mathrm{H^-}$, so the high-frequency dielectric $\varepsilon_{\infty}$ constant and thus the refractive index ($n \approx \sqrt{\varepsilon_{\infty}}$
) rises accordingly. The calculated high-frequency dielectric constants follow chemical intuition: fluorides are the least polarizable ($\varepsilon_{\infty}$ = $1.8$--$2.1$), chlorides are higher ($\varepsilon_{\infty}$ = $2.3$--$2.7$), and hydrides are the largest ($\varepsilon_{\infty}$ = $2.7$--$4.8$).

A consistent weakness remains in the prediction of the electronic band gap ($E_g$) using these functionals. For example, the calculated values for KF ($\sim 6\,\mathrm{eV}$) and KCl ($\sim 5\,\mathrm{eV}$) are significantly lower than their respective experimental measurements of $10.90\,\mathrm{eV}$ and $8.70\,\mathrm{eV}$ \cite{pantelides1975electronic}. It was previously found that other functionals, such as the modified Becke–Johnson (mBJ) exchange potential, can be used to obtain a more accurate estimate of the band gap \cite{tran2009accurate, camargo2012performance, messaoudi2015band}. The electronic band structures of fluoride, chloride and hydride materials can be found in Appendix B of the supplementary material, where the Fermi energy ($E_f$) is shifted to zero eV.

\begin{figure}[p]
  \centering
  \includegraphics[page=1,
                   width=\textwidth,
                   height=\textheight,
                   keepaspectratio]{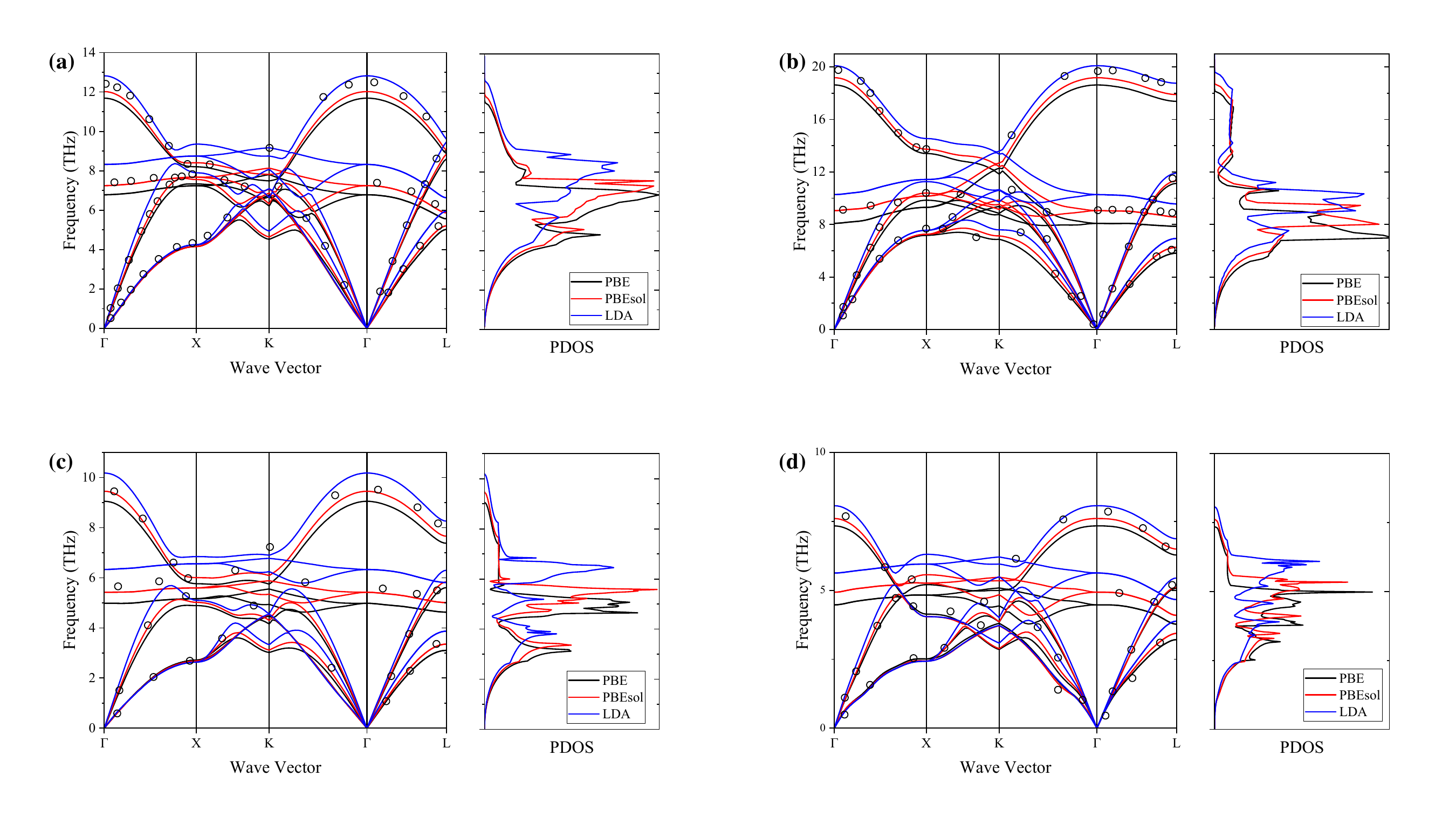}

  \includegraphics[page=2,
                   width=\textwidth,
                   height=\textheight,
                   keepaspectratio]{dispersions-all}

  \caption{Calculated phonon dispersion and phonon density of states for:
  (a) NaF, black circles are the measured data from Ref.~\cite{buyers1967lattice}.
  (b) LiF, black circles are the measured data from Ref.~\cite{dolling1968lattice}.
  (c) KF, black circles are the measured data from Ref.~\cite{buhrer1970lattice}.
  (d) NaCl, black circles are the measured data from Ref.~\cite{schmunk1970lattice}.
  (e) KCl, black circles are the measured data from Ref.~\cite{copley1969lattice}.
  (f) LiH, black circles are the measured data from Ref.~\cite{verble1968lattice}.
  (g) NaH. (h) KH.
  Calculations were done with PBE (black), PBEsol (red), and LDA (blue).}
  \label{fig:dispersion}
\end{figure}

The dispersion curves for the eight fluoride, chloride, and hydrides, seen in figure ~\ref{fig:dispersion}, are qualitatively consistent across PBE, PBEsol, and LDA and align with the experimental measurement results. The phonon frequency values of LO and TO modes at the gamma point are also given in table~\ref{tab:general-materials}. LDA leads to slightly higher phonon frequencies compared to PBE and PBEsol. Nevertheless, the predictions from all three functionals are in good agreement with experimental measurements along high-symmetry lines. A clear hierarchy emerges from the band shapes. Within each anion family, replacing Li with Na and then K progressively softens both acoustic and optical branches and lowers sound velocities. Across a fixed cation, moving from chloride to fluoride to hydride pushes the optical manifold upward and opens a pronounced acoustic–optical gap. The hydrides (LiH, NaH, KH) show the widest gap and the flattest high-frequency optic branches, reflecting the large mass contrast and strong long-range Coulomb coupling \cite{toberer2011phonon, driscoll2022long}. The large LO-TO splitting frequency values are due to strong polar phonon frequencies (creation of an oscillating electric field inside a crystal caused by phonons due to local charge polarization) \cite{bandyopadhyay2012physics}. The $\Gamma$-point LO--TO separations quantify these patterns. Using the tabulated PBE values, the hydrides have the largest splittings, with $\Delta_{\text{LO--TO}}$ of $14.87\,\mathrm{THz}$ for LiH, $11.78\,\mathrm{THz}$ for NaH, and $9.85\,\mathrm{THz}$ for KH. Fluorides are intermediate, with $10.54\,\mathrm{THz}$ for LiF, $4.89\,\mathrm{THz}$ for NaF, and $4.07\,\mathrm{THz}$ for KF. Chlorides are the smallest, with $2.87\,\mathrm{THz}$ for NaCl and $2.12\,\mathrm{THz}$ for KCl. The PDOS curves reflect these trends.

\newpage
\clearpage

\begin{table*}[htbp]
\scriptsize
\centering
\setlength{\tabcolsep}{1.5pt}
\renewcommand{\arraystretch}{0.8}

\caption{Lattice constant ($a$), high-frequency dielectric constant ($\varepsilon_\infty$), Born effective charge ($Z^*$), phonon frequencies at $\Gamma$ ($\omega$), and band gap ($E_g$) of calculated (PBE, PBEsol, LDA), other theoretical data, and experimental data for NaF, LiF, KF, NaCl, KCl, LiH, NaH and KH.}
\label{tab:general-materials}

\begin{tabularx}{\textwidth}{@{} l c c c >{\raggedright\arraybackslash}X >{\raggedright\arraybackslash}X @{}}
\hline
Property & PBE & PBEsol & LDA & Other theoretical data & Experiments \\
\hline

\multicolumn{6}{c}{\textbf{NaF}}\\
\hline
$a$ (\AA)&
4.72 & 4.65 & 4.52 &
4.67$^{\mathrm{a}}$, 4.69$^{\mathrm{b}}$, 4.63$^{\mathrm{c}}$, 4.80$^{\mathrm{d}}$, 4.70 (FP\,-LAPW)$^{\mathrm{e}}$, 4.69 (PP)$^{\mathrm{e}}$ &
4.61$^{\mathrm{f}}$, 4.63$^{\mathrm{g}}$ \\
$\varepsilon_\infty$ &
1.86 & 1.89 & 1.93 &
1.85$^{\mathrm{e}}$ &
1.75$^{\mathrm{h}}$ \\
$Z^*$ &
1.02 & 1.01 & 0.99 &
1.02$^{\mathrm{e}}$, 0.96$^{\mathrm{i}}$ &
1.02$^{\mathrm{j}}$ \\
$\omega_{\mathrm{LO}}$ (THz)&
11.67 & 12.01 & 12.81 &
11.90$^{\mathrm{e}}$ &
12.65$^{\mathrm{k}}$ \\
$\omega_{\mathrm{TO}}$ (THz)&
6.78 & 7.24 & 8.32 &
7.70$^{\mathrm{c}}$, 6.95$^{\mathrm{e}}$ &
7.86$^{\mathrm{h}}$, 7.51$^{\mathrm{k}}$ \\
$E_g$ (eV)&
6.02 & 6.11 & 6.34 &
12.00$^{\mathrm{a}}$, 6.12 (GGA)$^{\mathrm{e}}$, 7.11 (EV)$^{\mathrm{e}}$, 11.69 (mBJ)$^{\mathrm{e}}$, 5.94$^{\mathrm{l}}$ &
11.50$^{\mathrm{m}}$ \\

\hline
\multicolumn{6}{c}{\textbf{LiF}}\\
\hline
$a$ (\AA)&
4.07 & 4.01 & 3.91 &
4.02$^{\mathrm{c}}$, 3.89 (LDA)$^{\mathrm{n}}$, 4.01 (GGA)$^{\mathrm{n}}$, 4.00 (LDA)$^{\mathrm{o}}$, 4.10 (GGA)$^{\mathrm{o}}$, 4.07$^{\mathrm{p}}$ &
4.02$^{\mathrm{q}}$, 3.99$^{\mathrm{r}}$ \\
$\varepsilon_\infty$ &
2.04 & 2.08 & 2.11 &
2.04$^{\mathrm{o}}$, 2.19$^{\mathrm{s}}$ &
1.93$^{\mathrm{h}}$ \\
$Z^*$ &
1.05 & 1.05 & 1.04 &
0.99$^{\mathrm{i}}$, 1.03$^{\mathrm{o}}$, 1.06$^{\mathrm{p}}$ &
-- \\
$\omega_{\mathrm{LO}}$ (THz)&
18.63 & 19.18 & 20.09 &
19.89$^{\mathrm{o}}$, 19.92$^{\mathrm{p}}$ &
19.70$^{\mathrm{t}}$ \\
$\omega_{\mathrm{TO}}$ (THz)&
8.09 & 9.06 & 10.28 &
9.00$^{\mathrm{c}}$, 10.61$^{\mathrm{o}}$, 10.29$^{\mathrm{p}}$ &
9.54$^{\mathrm{h}}$, 9.15$^{\mathrm{t}}$ \\
$E_g$ (eV)&
8.82 & 9.04 & 9.64 &
9.11$^{\mathrm{l}}$, 8.73$^{\mathrm{p}}$, 7.64 (LDA)$^{\mathrm{u}}$, 9.15 (PBE)$^{\mathrm{u}}$, 8.70$^{\mathrm{v}}$ &
13.60$^{\mathrm{w}}$, 14.20$^{\mathrm{x}}$ \\

\hline
\multicolumn{6}{c}{\textbf{KF}}\\
\hline
$a$ (\AA)&
5.42 & 5.31 & 5.16 &
5.49$^{\mathrm{c}}$, 5.40$^{\mathrm{d}}$ &
5.31$^{\mathrm{f}}$, 5.29$^{\mathrm{r}}$, 5.35$^{\mathrm{y}}$ \\
$\varepsilon_\infty$ &
1.95 & 2.02 & 2.09 &
-- &
1.86$^{\mathrm{h}}$, 1.85$^{\mathrm{y}}$ \\
$Z^*$ &
1.13 & 1.14 & 1.15 &
0.92$^{\mathrm{z}}$ &
0.88$^{\mathrm{z}}$ \\
$\omega_{\mathrm{LO}}$ (THz)&
9.05 & 9.45 & 10.18 &
9.55$^{\mathrm{aa}}$ &
9.55$^{\mathrm{y}}$ \\
$\omega_{\mathrm{TO}}$ (THz)&
4.98 & 5.42 & 6.33 &
5.80$^{\mathrm{c}}$, 5.89$^{\mathrm{aa}}$ &
6.05$^{\mathrm{h}}$, 5.57$^{\mathrm{y}}$ \\
$E_g$ (eV)&
5.96 & 6.15 & 6.36 &
6.14$^{\mathrm{l}}$ &
10.90$^{\mathrm{ab}}$ \\

\hline
\multicolumn{6}{c}{\textbf{NaCl}}\\
\hline
$a$ (\AA)&
5.70 & 5.61 & 5.47 &
5.80$^{\mathrm{c}}$, 5.70$^{\mathrm{d}}$, 5.70 (FP\,-LAPW)$^{\mathrm{e}}$, 5.70 (PP)$^{\mathrm{e}}$,
5.52$^{\mathrm{ac}}$, 5.70$^{\mathrm{ad}}$, 5.47 (LDA)$^{\mathrm{ae}}$, 5.70 (GGA)$^{\mathrm{ae}}$ &
5.57$^{\mathrm{c}}$, 5.60$^{\mathrm{f}}$, 5.60$^{\mathrm{ac}}$, 5.60$^{\mathrm{ae}}$ \\
$\varepsilon_\infty$ &
2.48 & 2.57 & 2.67 &
2.47$^{\mathrm{e}}$, 2.55$^{\mathrm{af}}$ &
2.35$^{\mathrm{h}}$ \\
$Z^*$ &
1.11 & 1.09 & 1.07 &
1.11$^{\mathrm{e}}$, 0.99$^{\mathrm{i}}$, 1.10$^{\mathrm{af}}$ &
1.10$^{\mathrm{i}}$ \\
$\omega_{\mathrm{LO}}$ (THz)&
7.34 & 7.60 & 8.07 &
7.46$^{\mathrm{e}}$ &
7.93$^{\mathrm{ag}}$ \\
$\omega_{\mathrm{TO}}$ (THz)&
4.47 & 4.93 & 5.63 &
4.70$^{\mathrm{c}}$, 4.62$^{\mathrm{e}}$ &
5.34$^{\mathrm{h}}$, 5.17$^{\mathrm{ag}}$ \\
$E_g$ (eV)&
4.98 & 4.99 & 5.04 &
5.00$^{\mathrm{l}}$, 5.00 (GGA)$^{\mathrm{e}}$, 6.07 (EV)$^{\mathrm{e}}$, 8.41 (mBJ)$^{\mathrm{e}}$ &
8.75$^{\mathrm{ah}}$, 9.00$^{\mathrm{ai}}$ \\

\hline
\multicolumn{6}{c}{\textbf{KCl}}\\
\hline
$a$ (\AA)&
6.38 & 6.24 & 6.07 &
6.57$^{\mathrm{c}}$, 6.40$^{\mathrm{d}}$, 6.30$^{\mathrm{ac}}$ &
6.25$^{\mathrm{f}}$, 6.24$^{\mathrm{ac}}$ \\
$\varepsilon_\infty$ &
2.28 & 2.38 & 2.51 &
2.37$^{\mathrm{af}}$ &
2.20$^{\mathrm{h}}$, 2.15$^{\mathrm{aj}}$ \\
$Z^*$ &
1.13 & 1.12 & 1.11 &
1.13$^{\mathrm{af}}$ &
-- \\
$\omega_{\mathrm{LO}}$ (THz)&
5.89 & 6.16 & 6.66 &
6.34$^{\mathrm{aj}}$ &
-- \\
$\omega_{\mathrm{TO}}$ (THz)&
3.77 & 4.15 & 4.85 &
4.10$^{\mathrm{c}}$, 4.41$^{\mathrm{aj}}$ &
4.53$^{\mathrm{h}}$, 4.36$^{\mathrm{aj}}$ \\
$E_g$ (eV)&
5.05 & 5.11 & 5.12 &
5.32$^{\mathrm{l}}$, 5.50$^{\mathrm{ak}}$ &
8.70$^{\mathrm{ab}}$ \\

\hline
\multicolumn{6}{c}{\textbf{LiH}}\\
\hline
$a$ (\AA)&
4.01 & 3.98 & 3.91 &
3.89 (LDA)$^{\mathrm{n}}$, 3.95 (GGA)$^{\mathrm{n}}$, 4.02$^{\mathrm{al}}$, 4.00$^{\mathrm{am}}$ &
4.07$^{\mathrm{an}}$, 4.08$^{\mathrm{ao}}$, 4.08$^{\mathrm{ap}}$, 4.08$^{\mathrm{aq}}$ \\
$\varepsilon_\infty$ &
4.39 & 4.70 & 4.99 &
4.34$^{\mathrm{al}}$, 4.81$^{\mathrm{ar}}$ &
3.61$^{\mathrm{ao}}$ \\
$Z^*$ &
1.02 & 1.02 & 1.03 &
1.05$^{\mathrm{i}}$, 1.02$^{\mathrm{am}}$, 1.03$^{\mathrm{ar}}$ &
1.11$^{\mathrm{ar}}$, 0.99$^{\mathrm{as}}$ \\
$\omega_{\mathrm{LO}}$ (THz)&
32.43 & 32.37 & 32.96 &
28.30$^{\mathrm{ar}}$ &
32.21$^{\mathrm{an}}$, 33.50$^{\mathrm{ar}}$, 33.60$^{\mathrm{as}}$, 33.58$^{\mathrm{at}}$ \\
$\omega_{\mathrm{TO}}$ (THz)&
17.56 & 18.52 & 19.41 &
18.10$^{\mathrm{ar}}$, 17.86$^{\mathrm{au}}$ &
18.14$^{\mathrm{an}}$, 18.40$^{\mathrm{ar}}$, 17.76$^{\mathrm{as}}$, 17.70$^{\mathrm{at}}$ \\
$E_g$ (eV)&
2.98 & 2.73 & 2.54 &
3.00 (GGA)$^{\mathrm{al}}$, 2.61 (LDA)$^{\mathrm{al}}$, 4.75 (GW)$^{\mathrm{al}}$, 2.53$^{\mathrm{au}}$ &
5.00$^{\mathrm{av}}$, 4.92$^{\mathrm{aw}}$ \\
\hline
\multicolumn{6}{c}{\textbf{NaH}}\\
\hline
$a$ (\AA)&
4.84 & 4.79 & 4.69 &
4.83$^{\mathrm{al}}$, 4.83$^{\mathrm{am}}$, 4.82$^{\mathrm{ax}}$, 4.86$^{\mathrm{ay}}$,
4.77 (CN)$^{\mathrm{az}}$, 4.92 (QH)$^{\mathrm{az}}$,
4.87 (GGA)$^{\mathrm{ba}}$, 4.74 (LDA)$^{\mathrm{ba}}$ &
4.91$^{\mathrm{az}}$, 4.88$^{\mathrm{bb}}$ \\
$\varepsilon_\infty$ &
3.13 & 3.26 & 3.36 &
3.06$^{\mathrm{al}}$, 3.09$^{\mathrm{ax}}$,
3.12 (GGA)$^{\mathrm{ba}}$, 3.35 (LDA)$^{\mathrm{ba}}$ &
-- \\
$Z^*$ &
0.97 & 0.96 & 0.95 &
0.97$^{\mathrm{am}}$, 0.96$^{\mathrm{ax}}$,
0.99 (GGA)$^{\mathrm{ba}}$, 0.97 (LDA)$^{\mathrm{ba}}$ &
-- \\
$\omega_{\mathrm{LO}}$ (THz)&
26.75 & 26.65 & 27.52 &
27.06$^{\mathrm{am}}$ &
-- \\
$\omega_{\mathrm{TO}}$ (THz)&
14.97 & 15.44 & 16.83 &
16.60 (CN)$^{\mathrm{az}}$, 13.90 (QH)$^{\mathrm{az}}$ &
15.00$^{\mathrm{az}}$ \\
$E_g$ (eV)&
3.76 & 3.61 & 3.55 &
3.79 (GGA)$^{\mathrm{al}}$, 3.42 (LDA)$^{\mathrm{al}}$, 5.87 (GW)$^{\mathrm{al}}$, 4.90$^{\mathrm{ax}}$,
5.68 (GW)$^{\mathrm{bc}}$, 3.39 (LDA)$^{\mathrm{bc}}$ &
-- \\

\hline
\multicolumn{6}{c}{\textbf{KH}}\\
\hline
$a$ (\AA)&
5.70 & 5.60 & 5.47 &
5.84 (CN)$^{\mathrm{az}}$, 5.99 (QH)$^{\mathrm{az}}$,
5.69 (GGA)$^{\mathrm{ba}}$, 5.47 (LDA)$^{\mathrm{ba}}$ &
5.73$^{\mathrm{az}}$ \\
$\varepsilon_\infty$ &
2.72 & 2.84 & 2.97 &
2.69 (GGA)$^{\mathrm{ba}}$, 2.94 (LDA)$^{\mathrm{ba}}$ &
-- \\
$Z^*$ &
0.99 & 0.98 & 0.97 &
1.00 (GGA)$^{\mathrm{ba}}$, 0.98 (LDA)$^{\mathrm{ba}}$ &
-- \\
$\omega_{\mathrm{LO}}$ (THz)&
22.78 & 23.35 & 24.31 &
-- & -- \\
$\omega_{\mathrm{TO}}$ (THz)&
12.93 & 14.14 & 15.67 &
14.80 (CN)$^{\mathrm{az}}$, 12.60 (QH)$^{\mathrm{az}}$ &
-- \\
$E_g$ (eV)&
3.43 & 3.25 & 3.04 &
5.85 (GW)$^{\mathrm{bc}}$, 3.18 (LDA)$^{\mathrm{bc}}$ &
-- \\
\hline
\end{tabularx}

\begin{minipage}{\textwidth}
\footnotesize
$^{\mathrm{a}}$Ref.~\cite{wang2012phase}.
$^{\mathrm{b}}$Ref.~\cite{hartel2010relaxation}.
$^{\mathrm{c}}$Ref.~\cite{prencipe1995ab}.
$^{\mathrm{d}}$Ref.~\cite{solovyeva2016alchemical}.
$^{\mathrm{e}}$Ref.~\cite{messaoudi2015band}.
$^{\mathrm{f}}$Ref.~\cite{pies2006crystal}.
$^{\mathrm{g}}$Ref.~\cite{bruno2014crc}.
$^{\mathrm{h}}$Ref.~\cite{lowndes1969dielectric}.
$^{\mathrm{i}}$Ref.~\cite{shukla2000ab}.
$^{\mathrm{j}}$Ref.~\cite{born1996dynamical}.
$^{\mathrm{k}}$Ref.~\cite{buyers1967lattice}.
$^{\mathrm{l}}$Ref.~\cite{wu2017electronic}.
$^{\mathrm{m}}$Ref.~\cite{brown1970extreme}.
$^{\mathrm{n}}$Ref.~\cite{lindsay2016isotope}.
$^{\mathrm{o}}$Ref.~\cite{liang2018lattice}.
$^{\mathrm{p}}$Ref.~\cite{hou2016structural}.
$^{\mathrm{q}}$Ref.~\cite{ekinci2004thermal}.
$^{\mathrm{r}}$Ref.~\cite{wyckoff1966crystal}.
$^{\mathrm{s}}$Ref.~\cite{bernardini1998electronic}.
$^{\mathrm{t}}$Ref.~\cite{dolling1968lattice}.
$^{\mathrm{u}}$Ref.~\cite{guo2008ab}.
$^{\mathrm{v}}$Ref.~\cite{shirley1996detailed}.
$^{\mathrm{w}}$Ref.~\cite{roessler1967electronic}.
$^{\mathrm{x}}$Ref.~\cite{piacentini1976thermoreflectance}.
$^{\mathrm{y}}$Ref.~\cite{buhrer1970lattice}.
$^{\mathrm{z}}$Ref.~\cite{hass1963infrared}.
$^{\mathrm{aa}}$Ref.~\cite{cunningham1974second}.
$^{\mathrm{ab}}$Ref.~\cite{pantelides1975electronic}.
$^{\mathrm{ac}}$Ref.~\cite{froyen1986structural}.
$^{\mathrm{ad}}$Ref.~\cite{paier2006screened}.
$^{\mathrm{ae}}$Ref.~\cite{heyd2004efficient}.
$^{\mathrm{af}}$Ref.~\cite{togo2022mode}.
$^{\mathrm{ag}}$Ref.~\cite{raunio1969phonon}.
$^{\mathrm{ah}}$Ref.~\cite{miyata1968optical}.
$^{\mathrm{ai}}$Ref.~\cite{himpsel1978angle}.
$^{\mathrm{aj}}$Ref.~\cite{copley1969lattice}.
$^{\mathrm{ak}}$Ref.~\cite{erdinc2015ab}.
$^{\mathrm{al}}$Ref.~\cite{van2007electronic}.
$^{\mathrm{am}}$Ref.~\cite{yang2017thermal}.
$^{\mathrm{an}}$Ref.~\cite{verble1968lattice}.
$^{\mathrm{ao}}$Ref.~\cite{pretzel1960properties}.
$^{\mathrm{ap}}$Ref.~\cite{messer1960survey}.
$^{\mathrm{aq}}$Ref.~\cite{anderson1970isotopic}.
$^{\mathrm{ar}}$Ref.~\cite{blat1991calculations}.
$^{\mathrm{as}}$Ref.~\cite{brodsky1967infrared}.
$^{\mathrm{at}}$Ref.~\cite{laplaze1976etude}.
$^{\mathrm{au}}$Ref.~\cite{roma1996phonon}.
$^{\mathrm{av}}$Ref.~\cite{kondo1988effect}.
$^{\mathrm{aw}}$Ref.~\cite{plekhanov1998wannier}.
$^{\mathrm{ax}}$Ref.~\cite{ke2005decomposition}.
$^{\mathrm{ay}}$Ref.~\cite{sun2012structural}.
$^{\mathrm{az}}$Ref.~\cite{martins1990equations}.
$^{\mathrm{ba}}$Ref.~\cite{barrera2005lda}.
$^{\mathrm{bb}}$Ref.~\cite{humphries2016fluoride}.
$^{\mathrm{bc}}$Ref.~\cite{lebegue2003implementation}.
\end{minipage}

\end{table*}

\newpage
\clearpage

\subsection{Mechanical Properties}

The mechanical properties of fluorides, chlorides and hydrides, including the bulk modulus and elastic constants, are summarized in table~\ref{tab:mechanical-materials}. For cubic crystals, the three independent elastic constants are $C_{11}$, $C_{12}$, and $C_{44}$. Our calculated bulk modulus and elastic constant values are in good agreement with the experimental values. We computed the bulk modulus $B$ and the elastic constants $C_{ij}$ at $0\,\mathrm{K}$. LDA typically yields the largest bulk moduli, PBEsol is intermediate, and PBE gives the smallest (e.g., LiF: $B_{\mathrm{PBE}} = 67.17\,\mathrm{GPa}$, $B_{\mathrm{PBEsol}} = 72.55\,\mathrm{GPa}$, $B_{\mathrm{LDA}} = 86.78\,\mathrm{GPa}$)
. This ordering follows from their equilibrium volumes: LDA overbinds (smaller lattice constants), which increases the curvature of the energy--volume relation \(B_0 = V_0\,\left.\mathrm{d}^2E/\mathrm{d}V^2\right|_{V_0}\) \cite{rodney2005discrete}; PBE underbinds (larger volumes), which reduces that curvature, and PBEsol was tuned for solids and therefore falls between. Although $B_0$ includes the factor $V_0$, the functional trend is governed primarily by the stiffness of the energy--volume curve at equilibrium: LDA overbinding makes the $E(V)$ minimum noticeably narrower (larger curvature), and this increase in curvature more than offsets the smaller $V_0$, resulting in systematically larger bulk moduli. Replacing Li with Na and then K systematically softens the lattice at a fixed anion. For fluorides, using PBE for comparisons, the bulk modulus decreases from
$B=67.17\,\mathrm{GPa}$ (LiF) to $46.79\,\mathrm{GPa}$ (NaF) and $29.01\,\mathrm{GPa}$ (KF). For chlorides, it decreases from $23.67\,\mathrm{GPa}$ (NaCl) to $16.27\,\mathrm{GPa}$ (KCl). For hydrides, it decreases from $34.24\,\mathrm{GPa}$ (LiH) to $22.16\,\mathrm{GPa}$ (NaH) and $13.74\,\mathrm{GPa}$ (KH). In magnitude, fluorides are the stiffest for a given cation, chlorides are intermediate, and hydrides are the softest for Na and K.

For cubic crystals, mechanical stability requires the elastic
constants to satisfy four conditions: $C_{11}>0$, $C_{44}>0$, $C_{11}-C_{12}>0$, and $C_{11}+2C_{12}>0$ \cite{hou2016structural}. The calculated values of the elastic constants for the fluoride, chloride, and hydride materials satisfy all these conditions. Comparing theoretical elastic constants with experiments is challenging because of the uncertainties in experimentally determining $C_{12}$, which is a linear combination of the elastic constants in the <110> direction, whereas $C_{11}$ and $C_{44}$ values can be measured directly via ultrasonic techniques\cite{cohen1975theory}. Directional elasticity is captured by the Zener anisotropy factor $AF = 2C_{44}/(C_{11}-C_{12})$, where $AF = 1$ for an elastic isotropic medium \cite{zener1949elasticity}. Using PBE elastic constants, the Zener anisotropy factors are: LiF $AF\!\approx\!1.90$, LiH $AF\!\approx\!1.65$, and NaH $AF\!\approx\!1.45$ are anisotropic;
KH $AF\!\approx\!0.94$ is nearly isotropic;
NaF $AF\!\approx\!0.72$, NaCl $AF\!\approx\!0.69$, KF $AF\!\approx\!0.61$, and KCl $AF\!\approx\!0.44$ are isotropic. This matches the $AF$ values reported by Sirdeshmukh \textit{et al} \cite{sirdeshmukh2001alkali}.

From the elastic constants, one can also inspect the data to determine whether the material is brittle or ductile. According to Cauchy pressure ($C_{12}$ – $C_{44}$), if the difference is positive, the material is ductile; otherwise, it is brittle \cite{pettifor1991bonding}. Based on this, all eight materials are brittle (i.e., negative difference).

\begin{table*}[htbp]
\scriptsize
\centering

\caption{Bulk modulus ($B$) and elastic constants ($C_{ij}$)of calculated (PBE, PBEsol, LDA), other theoretical data, and experimental data for NaF, LiF, KF, NaCl, KCl, LiH, NaH and KH.}
\label{tab:mechanical-materials}

\begin{tabularx}{\textwidth}{@{} l c c c >{\raggedright\arraybackslash}X >{\raggedright\arraybackslash}X @{}}
\hline
Property & PBE & PBEsol & LDA & Other theoretical data & Experiments \\
\hline

\multicolumn{6}{c}{\textbf{NaF}}\\
\hline
$B$ (GPa)&
46.79 & 48.46 & 61.14 &
48.50$^{\mathrm{a}}$, 40.80$^{\mathrm{b}}$, 61.60$^{\mathrm{c}}$, 42.50$^{\mathrm{d}}$, 51.10$^{\mathrm{e}}$ &
45.70$^{\mathrm{f}}$, 53.80$^{\mathrm{g}}$, 51.40$^{\mathrm{h}}$ \\
$C_{11}$ (GPa)&
95.68 & 101.86 & 133.97 &
77.20$^{\mathrm{d}}$, 107.80$^{\mathrm{e}}$, 145.14$^{\mathrm{i}}$ &
97.00$^{\mathrm{f}}$, 115.40$^{\mathrm{g}}$, 108.50$^{\mathrm{h}}$ \\
$C_{12}$ (GPa)&
22.84 & 21.75 & 24.73 &
25.20$^{\mathrm{d}}$, 22.80$^{\mathrm{e}}$, 33.83$^{\mathrm{i}}$ &
23.80$^{\mathrm{f}}$, 23.00$^{\mathrm{g}}$, 22.90$^{\mathrm{h}}$ \\
$C_{44}$ (GPa)&
26.35 & 26.59 & 28.31 &
25.70$^{\mathrm{d}}$, 33.60$^{\mathrm{e}}$, 33.83$^{\mathrm{i}}$ &
28.22$^{\mathrm{f}}$, 29.80$^{\mathrm{g}}$, 28.99$^{\mathrm{h}}$ \\

\hline
\multicolumn{6}{c}{\textbf{LiF}}\\
\hline
$B$ (GPa)&
67.17 & 72.55 & 86.78 &
87.11$^{\mathrm{c}}$, 75.90$^{\mathrm{e}}$, 67.71 (GGA)$^{\mathrm{j}}$, 73.59 (LDA)$^{\mathrm{j}}$, 72.99 (Birch--Murnaghan EOS)$^{\mathrm{k}}$, 70.53 (Vinet EOS)$^{\mathrm{k}}$ &
66.40$^{\mathrm{f}}$, 76.90$^{\mathrm{g}}$ \\
$C_{11}$ (GPa)&
109.45 & 126.71 & 143.12 &
125.80$^{\mathrm{e}}$, 224.83$^{\mathrm{i}}$, 153.15$^{\mathrm{k}}$ &
113.73$^{\mathrm{f}}$, 135.80$^{\mathrm{g}}$ \\
$C_{12}$ (GPa)&
46.03 & 45.48 & 50.61 &
50.90$^{\mathrm{e}}$, 121.09$^{\mathrm{i}}$, 45.62$^{\mathrm{k}}$ &
47.59$^{\mathrm{f}}$, 47.40$^{\mathrm{g}}$ \\
$C_{44}$ (GPa)&
60.32 & 61.91 & 68.07 &
76.10$^{\mathrm{e}}$, 121.09$^{\mathrm{i}}$, 55.08$^{\mathrm{k}}$ &
63.68$^{\mathrm{f}}$, 68.70$^{\mathrm{g}}$ \\

\hline
\multicolumn{6}{c}{\textbf{KF}}\\
\hline
$B$ (GPa)&
29.01 & 33.21 & 43.81 &
29.80$^{\mathrm{b}}$, 29.70$^{\mathrm{e}}$ &
35.50$^{\mathrm{g}}$, 34.20$^{\mathrm{h}}$ \\
$C_{11}$ (GPa)&
58.87 & 69.90 & 96.63 &
66.10$^{\mathrm{e}}$, 84.76$^{\mathrm{i}}$ &
79.90$^{\mathrm{g}}$, 75.70$^{\mathrm{h}}$ \\
$C_{12}$ (GPa)&
11.08 & 12.86 & 13.40 &
11.90$^{\mathrm{e}}$, 12.83$^{\mathrm{i}}$ &
13.40$^{\mathrm{g}}$, 13.50$^{\mathrm{h}}$ \\
$C_{44}$ (GPa)&
13.65 & 13.62 & 13.54 &
15.60$^{\mathrm{e}}$, 12.83$^{\mathrm{i}}$ &
13.30$^{\mathrm{g}}$, 13.36$^{\mathrm{h}}$ \\

\hline
\multicolumn{6}{c}{\textbf{NaCl}}\\
\hline
$B$ (GPa)&
23.67 & 25.82 & 32.06 &
22.90$^{\mathrm{b}}$, 32.20$^{\mathrm{c}}$, 22.30$^{\mathrm{e}}$, 31.20$^{\mathrm{l}}$ &
28.60$^{\mathrm{g}}$, 26.60$^{\mathrm{h}}$ \\
$C_{11}$ (GPa)&
47.35 & 54.85 & 70.28 &
47.40$^{\mathrm{e}}$, 72.44$^{\mathrm{i}}$ &
61.10$^{\mathrm{g}}$, 57.33$^{\mathrm{h}}$ \\
$C_{12}$ (GPa)&
11.83 & 11.31 & 12.95 &
9.80$^{\mathrm{e}}$, 19.01$^{\mathrm{i}}$ &
12.30$^{\mathrm{g}}$, 11.23$^{\mathrm{h}}$ \\
$C_{44}$ (GPa)&
12.24 & 11.92 & 12.71 &
15.30$^{\mathrm{e}}$, 19.01$^{\mathrm{i}}$ &
13.60$^{\mathrm{g}}$, 13.31$^{\mathrm{h}}$ \\

\hline
\multicolumn{6}{c}{\textbf{KCl}}\\
\hline
$B$ (GPa)&
16.27 & 18.47 & 24.32 &
16.10$^{\mathrm{b}}$, 24.28$^{\mathrm{c}}$, 15.70$^{\mathrm{e}}$, 24.80$^{\mathrm{l}}$ &
20.80$^{\mathrm{g}}$ \\
$C_{11}$ (GPa)&
36.19 & 43.16 & 59.51 &
36.30$^{\mathrm{e}}$, 54.53$^{\mathrm{i}}$, 47.20$^{\mathrm{m}}$ &
50.90$^{\mathrm{g}}$, 46.00$^{\mathrm{m}}$ \\
$C_{12}$ (GPa)&
6.31 & 6.12 & 6.73 &
5.40$^{\mathrm{e}}$, 8.12$^{\mathrm{i}}$, 6.30$^{\mathrm{m}}$ &
5.80$^{\mathrm{g}}$, 5.80$^{\mathrm{m}}$ \\
$C_{44}$ (GPa)&
6.62 & 6.46 & 6.43 &
7.60$^{\mathrm{e}}$, 8.12$^{\mathrm{i}}$, 7.20$^{\mathrm{m}}$ &
6.70$^{\mathrm{g}}$, 6.53$^{\mathrm{m}}$ \\

\hline
\multicolumn{6}{c}{\textbf{LiH}}\\
\hline
$B$ (GPa)&
34.24 & 37.06 & 40.60 &
34.03 (GGA)$^{\mathrm{n}}$, 41.33 (LDA)$^{\mathrm{n}}$, 36.07$^{\mathrm{o}}$, 36.20 (GGA)$^{\mathrm{p}}$, 40.50 (LDA)$^{\mathrm{p}}$, 34.30$^{\mathrm{q}}$, 40.00$^{\mathrm{r}}$ &
32.00$^{\mathrm{r}}$, 32.35$^{\mathrm{s}}$ \\
$C_{11}$ (GPa)&
73.06 & 85.34 & 95.36 &
100.90$^{\mathrm{p}}$, 82.70$^{\mathrm{q}}$, 79.80$^{\mathrm{t}}$, 78.90$^{\mathrm{u}}$, 65.82$^{\mathrm{v}}$ &
99.10$^{\mathrm{q}}$, 67.20$^{\mathrm{s}}$, 74.10$^{\mathrm{t}}$ \\
$C_{12}$ (GPa)&
14.03 & 10.42 & 10.22 &
10.30$^{\mathrm{p}}$, 10.70$^{\mathrm{q}}$, 45.90$^{\mathrm{t}}$, 44.30$^{\mathrm{u}}$, 15.44$^{\mathrm{v}}$ &
11.20$^{\mathrm{q}}$, 14.93$^{\mathrm{s}}$, 14.20$^{\mathrm{t}}$ \\
$C_{44}$ (GPa)&
48.62 & 48.31 & 51.48 &
51.40$^{\mathrm{p}}$, 52.50$^{\mathrm{q}}$, 45.70$^{\mathrm{t}}$, 44.10$^{\mathrm{u}}$, 42.91$^{\mathrm{v}}$ &
59.80$^{\mathrm{q}}$, 46.37$^{\mathrm{s}}$, 48.40$^{\mathrm{t}}$ \\

\hline
\multicolumn{6}{c}{\textbf{NaH}}\\
\hline
$B$ (GPa)&
22.16 & 23.92 & 27.43 &
24.00 (GGA)$^{\mathrm{p}}$, 27.20 (LDA)$^{\mathrm{p}}$, 21.60$^{\mathrm{q}}$, 28.00$^{\mathrm{r}}$, 22.90$^{\mathrm{w}}$, 22.93$^{\mathrm{x}}$, 23.70$^{\mathrm{y}}$ &
19.00$^{\mathrm{r}}$, 19.40\,$\pm$\,2.00$^{\mathrm{z}}$, 14.30\,$\pm$\,1.50$^{\mathrm{aa}}$ \\
$C_{11}$ (GPa)&
43.69 & 47.19 & 55.25 &
56.90$^{\mathrm{p}}$, 53.20$^{\mathrm{q}}$, 47.30$^{\mathrm{u}}$, 42.12$^{\mathrm{x}}$ &
-- \\
$C_{12}$ (GPa)&
12.89 & 12.29 & 13.52 &
15.90$^{\mathrm{p}}$, 14.80$^{\mathrm{q}}$, 22.50$^{\mathrm{u}}$, 13.06$^{\mathrm{x}}$ &
-- \\
$C_{44}$ (GPa)&
22.35 & 22.31 & 24.19 &
24.70$^{\mathrm{p}}$, 22.70$^{\mathrm{q}}$, 22.50$^{\mathrm{u}}$, 22.02$^{\mathrm{x}}$ &
-- \\

\hline
\multicolumn{6}{c}{\textbf{KH}}\\
\hline
$B$ (GPa)&
13.74 & 14.67 & 17.40 &
13.40 (GGA)$^{\mathrm{p}}$, 17.40 (LDA)$^{\mathrm{p}}$, 13.30$^{\mathrm{q}}$, 19.00$^{\mathrm{r}}$ &
16.00$^{\mathrm{r}}$, 15.60\,$\pm$\,1.50$^{\mathrm{aa}}$ \\
$C_{11}$ (GPa)&
28.32 & 31.81 & 39.50 &
39.40$^{\mathrm{p}}$, 26.80$^{\mathrm{q}}$ &
-- \\
$C_{12}$ (GPa)&
6.45 & 6.11 & 6.35 &
6.30$^{\mathrm{p}}$, 6.50$^{\mathrm{q}}$ &
-- \\
$C_{44}$ (GPa)&
10.32 & 10.09 & 10.43 &
10.40$^{\mathrm{p}}$, 10.60$^{\mathrm{q}}$ &
-- \\
\hline
\end{tabularx}

\begin{minipage}{\textwidth}
\footnotesize
$^{\mathrm{a}}$Ref.~\cite{wang2012phase}.
$^{\mathrm{b}}$Ref.~\cite{solovyeva2016alchemical}.
$^{\mathrm{c}}$Ref.~\cite{otero2011gibbs2}.
$^{\mathrm{d}}$Ref.~\cite{recio1990theoretical}.
$^{\mathrm{e}}$Ref.~\cite{prencipe1995ab}.
$^{\mathrm{f}}$Ref.~\cite{miller1964pressure}.
$^{\mathrm{g}}$Ref.~\cite{haussuhl1960thermo} (data taken from Ref.~\cite{prencipe1995ab}).
$^{\mathrm{h}}$Ref.~\cite{lewis1967elastic} (bulk modulus values from Ref.~\cite{solovyeva2016alchemical}, measured at $T=4.2$\,K).
$^{\mathrm{i}}$Ref.~\cite{sangster1978interionic}.
$^{\mathrm{j}}$Ref.~\cite{liang2018lattice}.
$^{\mathrm{k}}$Ref.~\cite{hou2016structural}.
$^{\mathrm{l}}$Ref.~\cite{froyen1986structural}.
$^{\mathrm{m}}$Ref.~\cite{copley1969lattice}.
$^{\mathrm{n}}$Ref.~\cite{zhang2007first}.
$^{\mathrm{o}}$Ref.~\cite{bouhadda2007first}.
$^{\mathrm{p}}$Ref.~\cite{barrera2005lda}.
$^{\mathrm{q}}$Ref.~\cite{zhang2007phonon}.
$^{\mathrm{r}}$Ref.~\cite{rodriguez1992pressure}.
$^{\mathrm{s}}$Ref.~\cite{gerlich1974pressure}.
$^{\mathrm{t}}$Ref.~\cite{pandey1985intrinsic}.
$^{\mathrm{u}}$Ref.~\cite{haque1990hydrides}.
$^{\mathrm{v}}$Ref.~\cite{dyck1981lattice}.
$^{\mathrm{w}}$Ref.~\cite{sun2011ab}.
$^{\mathrm{x}}$Ref.~\cite{sun2012structural}.
$^{\mathrm{y}}$Ref.~\cite{ojwang2008modeling}.
$^{\mathrm{z}}$Ref.~\cite{duclos1987high}.
$^{\mathrm{aa}}$Ref.~\cite{hochheimer1985high}.
\end{minipage}

\end{table*}

\newpage
\clearpage

\subsection{Thermal Properties}

We calculate the room temperature lattice thermal conductivity, specific heat, average Gr\"uneisen parameter, and density values at room temperature. The computed values are listed in table~\ref{tab:thermal-materials} and compared with theoretical and experimental values from the literature. PBEsol generally yields thermal conductivities \(k\) that lie between those predicted by PBE (lower) and LDA (higher). This ordering is consistent with LDA’s stronger binding (smaller equilibrium volume), which typically stiffens the phonon spectrum and increases phonon transport (via higher group velocities and longer lifetimes), whereas PBE’s weaker binding shifts \(k\) downward, with PBEsol remaining intermediate. Some examples include: for NaF, PBE (18.50 W m$^{-1}$ K$^{-1}$) matches experiment (18.40, 18.70) \cite{morelli2006high, smirnov1845thermal}, while LDA overestimates (31.50); for KCl, PBEsol (6.90 W m$^{-1} $K$^{-1}$) tracks experiments (7.10, 6.70) \cite{morelli2006high, sirdeshmukh2001alkali}; for KF and NaCl, LDA (7.32\ \text{and}\ 7.86 W m$^{-1}$ K$^{-1}$) is the closest to the measured values (7.10) \cite{morelli2006high, sirdeshmukh2001alkali}. This variation indicates that the best agreement is material-dependent, because each functional shifts the predicted lattice constant and phonon spectrum differently, which in turn changes both phonon velocities and the strength of anharmonic scattering and, consequently, impacts the thermal conductivity. Specific heats are generally in good agreement with the experimental values across functionals (e.g., for NaCl \(c_p \approx 0.82\text{--}0.83~\mathrm{kJ\,kg^{-1}\,K^{-1}}\) vs. \(0.86 ~\mathrm{kJ\,kg^{-1}\,K^{-1}}\) \cite{haynes2016crc}). This limited XC sensitivity is expected because $c_p$ is governed primarily by the overall vibrational degrees of freedom and is therefore only weakly affected by modest, functional-induced shifts in the phonon spectrum. Average Gr\"uneisen parameters fall within 1.4–2.1 for all materials. On the other hand, density orders across the series (fluorides highest, chlorides intermediate, hydrides lowest), and the computed values closely track the experimental results (PBEsol performs best for halides, while PBE performs best for hydrides). For instance, LiF, PBEsol predicts a value of 2.67 g cm$^{-3}$, closely matching experiment (2.639) \cite{miller1964pressure}, whereas KH, PBE predicts a value of  1.44 g cm$^{-3}$, closely matching experiments (1.43) \cite{haynes2016crc, mueller2013metal}.

\begin{figure}[p]
  \centering
  \includegraphics[page=1,
                   width=\textwidth,
                   height=0.4\textheight,
                   keepaspectratio]{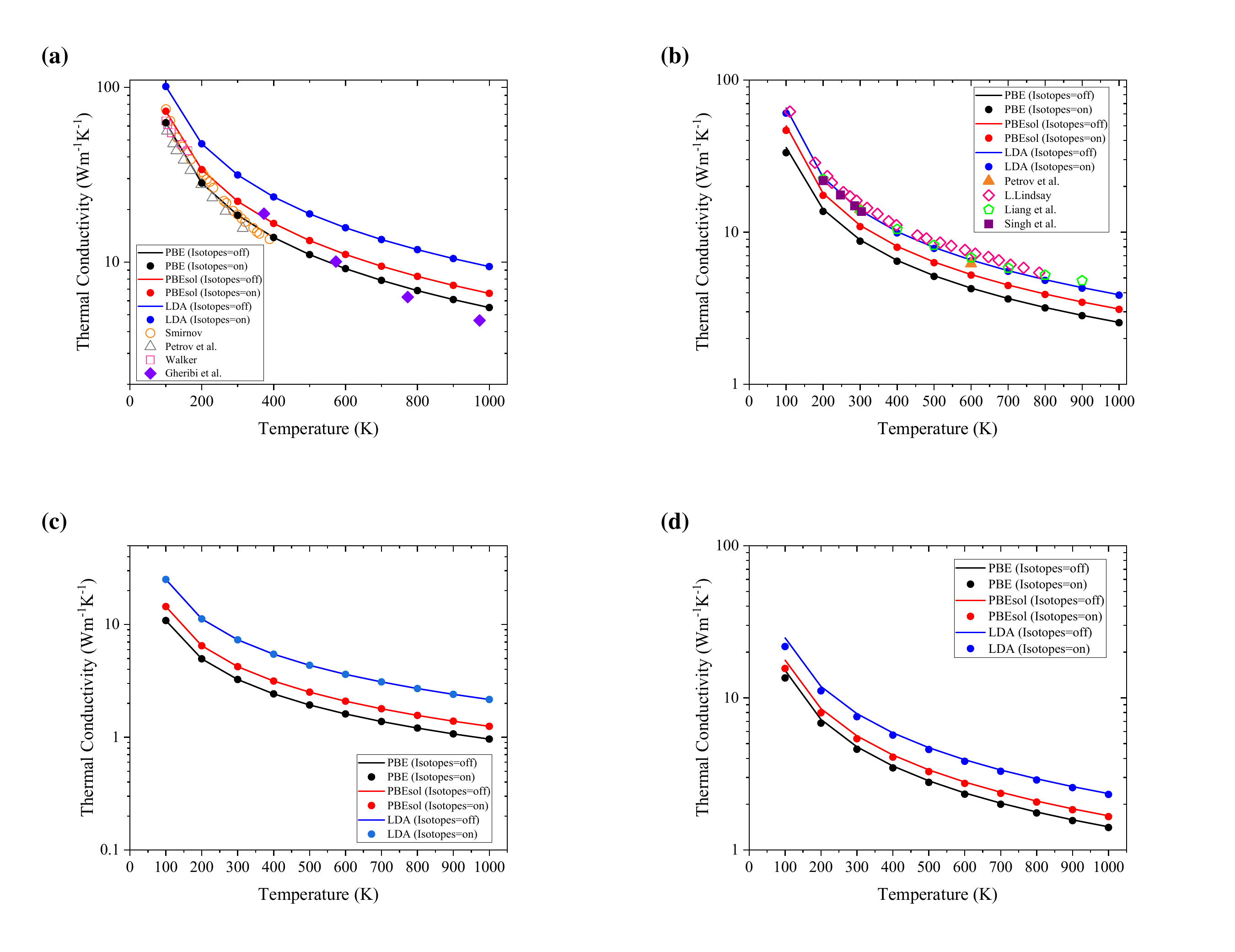}

  \includegraphics[page=2,
                   width=\textwidth,
                   height=0.4\textheight,
                   keepaspectratio]{thermal_all}
  \caption{
  Calculated lattice thermal conductivity for: (a) NaF. (b) LiF. (c) KF. (d) NaCl.
  (e) KCl. (f) LiH. (g) NaH. (h) KH. Calculations were done with PBE (black solid line),
  PBEsol (red solid line), and LDA (blue solid line). Calculations also show the effect
  of the isotopes on the thermal conductivity. Experimental data:
  Ref.~\cite{smirnov1845thermal} (orange circles),
  Ref.~\cite{petrov1974temperature} (gray triangles),
  Ref.~\cite{walker1963thermal} (pink squares),
  Ref.~\cite{petrov1976behaviour} (orange triangle),
  Ref.~\cite{vetrano1957batelle} (pink triangles),
  Ref.~\cite{slack1973nonmetallic} (orange squares).
  Theoretical data: Ref.~\cite{gheribi2016formulation} (purple diamonds),
  Ref.~\cite{lindsay2016isotope} (pink diamonds),
  Ref.~\cite{liang2018lattice} (green pentagons),
  Ref.~\cite{singh2003effects} (purple squares),
  Ref.~\cite{yang2017thermal} (purple pentagons).}
  \label{fig:thermal}
\end{figure}

The lattice thermal conductivity is calculated for temperatures ranging from 100 K to 1000 K (see figure ~\ref{fig:thermal}) and compared with the available data from the literature. The effect of isotopes on the thermal conductivity calculations is also shown in the same figure. All compounds show the expected monotonic decrease of $k$ with $T$. Including naturally abundant isotope scattering in the calculations mainly depresses $k$ (below 200 K) and has less impact at high $T$, where the three functionals produce nearly parallel slopes. Isotope scattering can significantly impact the lattice thermal conductivity $k$ in materials with mixed natural isotopes, but has little to no effect when the constituent elements are effectively monoisotopic. For example, NaF shows essentially no isotope effect because the natural abundance of Na is \(\sim 100\%\) \(^{23}\mathrm{Na}\) and the natural abundance of F is \(\sim 100\%\) \(^{19}\mathrm{F}\), whereas LiF exhibits a reduction in \(k\) when isotopes are included due to its mixed lithium composition (\(\sim 92.5\%\) \(^{7}\mathrm{Li}\), \(\sim 7.5\%\) \(^{6}\mathrm{Li}\) \cite{tomascak2004developments}) that introduces phonon mass-disorder scattering most evident at low $T$.

\begin{table*}[htbp]
\scriptsize
\centering
\setlength{\tabcolsep}{1.5pt}
\renewcommand{\arraystretch}{1.0}

\caption{Thermal conductivity ($k$), specific heat ($c_p$), average Gr\"uneisen parameter ($\bar{\gamma}$), and density ($\rho$) of calculated (PBE, PBEsol, LDA), other theoretical data, and experimental data for NaF, LiF, KF, NaCl, KCl, LiH, NaH and KH at room temperature.}
\label{tab:thermal-materials}

\begin{tabularx}{\textwidth}{@{} l c c c >{\raggedright\arraybackslash}X >{\raggedright\arraybackslash}X @{}}
\hline
Property & PBE & PBEsol & LDA & Other theoretical data & Experiments \\
\hline

\multicolumn{6}{c}{\textbf{NaF}}\\
\hline
$k$ (\(\mathrm{W\,m^{-1}\,K^{-1}}\))&
18.54  & 22.26  & 31.48 &
17.90$^{\mathrm{a}}$, 4.52$^{\mathrm{b}}$ &
18.40$^{\mathrm{a}}$, 18.70$^{\mathrm{c}}$ \\
$c_p$ (\(\mathrm{kJ\,kg^{-1}\,K^{-1}}\))&
1.09 & 1.08  & 1.06 &
-- &
1.09$^{\mathrm{d}}$ \\
$\bar{\gamma}$ &
1.78  & 1.61 & 1.59 &
2.20$^{\mathrm{b}}$, 1.79$^{\mathrm{e}}$ &
1.50$^{\mathrm{a}}$, 1.83$^{\mathrm{e}}$ \\
$\rho$ (\(\mathrm{g\,cm^{-3}}\))&
2.66 & 2.78 & 3.01 &
2.80$^{\mathrm{f}}$ &
2.80$^{\mathrm{g}}$, 2.80$^{\mathrm{h}}$, 2.80$^{\mathrm{i}}$ \\

\hline
\multicolumn{6}{c}{\textbf{LiF}}\\
\hline
$k$ (\(\mathrm{W\,m^{-1}\,K^{-1}}\))&
8.88 & 11.07 & 13.91 &
19.40$^{\mathrm{a}}$, 8.71$^{\mathrm{b}}$, 22.42 ($a_0=3.89$\,\AA)$^{\mathrm{j}}$, 14.59 ($a(300\,\mathrm{K})=4.00$\,\AA)$^{\mathrm{j}}$, 13.89 ($a_0=4.00$\,\AA)$^{\mathrm{k}}$, 13.48 ($a(300\,\mathrm{K})=4.03$\,\AA)$^{\mathrm{k}}$ &
17.60$^{\mathrm{a}}$, 14.09$^{\mathrm{l}}$, 14.20$^{\mathrm{m}}$ \\
$c_p$ (\(\mathrm{kJ\,kg^{-1}\,K^{-1}}\))&
1.62 & 1.59 & 1.55 &
1.62$^{\mathrm{d}}$ &
1.54$^{\mathrm{d}}$, 1.62$^{\mathrm{m}}$ \\
$\bar{\gamma}$ &
2.08 & 1.88 & 1.54 &
2.02$^{\mathrm{b}}$, 1.48$^{\mathrm{e}}$, 1.24 ($a_0=4.00$\,\AA)$^{\mathrm{k}}$, 1.36 ($a(300\,\mathrm{K})=4.03$\,\AA)$^{\mathrm{k}}$ &
1.50$^{\mathrm{a}}$, 1.64$^{\mathrm{e}}$ \\
$\rho$ (\(\mathrm{g\,cm^{-3}}\))&
2.56 & 2.67 & 2.88 &
2.64$^{\mathrm{f}}$ &
2.64$^{\mathrm{h}}$, 2.64$^{\mathrm{m}}$ \\

\hline
\multicolumn{6}{c}{\textbf{KF}}\\
\hline
$k$ (\(\mathrm{W\,m^{-1}\,K^{-1}}\))&
3.25 & 4.23 & 7.32 &
5.80$^{\mathrm{a}}$, 2.68$^{\mathrm{b}}$ &
7.10$^{\mathrm{m}}$ \\
$c_p$ (\(\mathrm{kJ\,kg^{-1}\,K^{-1}}\))&
0.82 & 0.82 & 0.81 &
-- &
0.84$^{\mathrm{m}}$ \\
$\bar{\gamma}$ &
1.87 & 1.85 & 1.67 &
2.29$^{\mathrm{b}}$, 1.88$^{\mathrm{e}}$ &
1.52$^{\mathrm{a}}$, 1.58$^{\mathrm{e}}$ \\
$\rho$ (\(\mathrm{g\,cm^{-3}}\))&
2.42 & 2.58 & 2.81 &
2.53$^{\mathrm{f}}$ &
2.48$^{\mathrm{g}}$, 2.48$^{\mathrm{i}}$, 2.48$^{\mathrm{m}}$ \\

\hline
\multicolumn{6}{c}{\textbf{NaCl}}\\
\hline
$k$ (\(\mathrm{W\,m^{-1}\,K^{-1}}\))&
4.77 & 5.62 & 7.86 &
4.80$^{\mathrm{a}}$, 2.43$^{\mathrm{b}}$ &
7.10$^{\mathrm{a}}$, 6.32$^{\mathrm{m}}$ \\
$c_p$ (\(\mathrm{kJ\,kg^{-1}\,K^{-1}}\))&
0.83 & 0.82 & 0.82 &
-- &
0.86$^{\mathrm{i}}$ \\
$\bar{\gamma}$ &
2.04 & 1.96 & 1.83 &
2.23$^{\mathrm{b}}$, 1.83$^{\mathrm{e}}$ &
1.56$^{\mathrm{a}}$, 1.64$^{\mathrm{e}}$ \\
$\rho$ (\(\mathrm{g\,cm^{-3}}\))&
2.09 & 2.20 & 2.37 &
2.16$^{\mathrm{f}}$ &
2.16$^{\mathrm{g}}$, 2.16$^{\mathrm{i}}$, 2.17$^{\mathrm{m}}$ \\

\hline
\multicolumn{6}{c}{\textbf{KCl}}\\
\hline
$k$ (\(\mathrm{W\,m^{-1}\,K^{-1}}\))&
4.80 & 6.90 & 10.60 &
3.80$^{\mathrm{a}}$, 1.40$^{\mathrm{b}}$ &
7.10$^{\mathrm{a}}$, 6.70$^{\mathrm{m}}$ \\
$c_p$ (\(\mathrm{kJ\,kg^{-1}\,K^{-1}}\))&
0.65 & 0.65 & 0.65 &
-- &
0.69$^{\mathrm{i}}$, 0.69$^{\mathrm{m}}$ \\
$\bar{\gamma}$ &
1.78 & 1.79 & 1.61 &
2.38$^{\mathrm{b}}$, 2.12$^{\mathrm{e}}$ &
1.45$^{\mathrm{a}}$, 1.49$^{\mathrm{e}}$ \\
$\rho$ (\(\mathrm{g\,cm^{-3}}\))&
1.91 & 2.04 & 2.21 &
1.99$^{\mathrm{f}}$ &
1.98$^{\mathrm{i}}$, 1.99$^{\mathrm{m}}$ \\

\hline
\multicolumn{6}{c}{\textbf{LiH}}\\
\hline
$k^\ast$ (\(\mathrm{W\,m^{-1}\,K^{-1}}\))&
12.47 & 13.76 & 16.98 &
25.51 ($a_0=3.89$\,\AA)$^{\mathrm{j}}$, 23.00 (300\,K)$^{\mathrm{j}}$, 12.98$^{\mathrm{n}}$ &
12.47$^{\mathrm{o}}$ \\
$c_p$ (\(\mathrm{kJ\,kg^{-1}\,K^{-1}}\))&
3.69 & 3.57 & 3.45 &
-- &
3.64$^{\mathrm{p}}$, 3.79$^{\mathrm{q}}$ \\
$\bar{\gamma}$ &
1.64 & 1.57 & 1.42 &
1.64$^{\mathrm{r}}$ &
1.28$^{\mathrm{s}}$ \\
$\rho$ (\(\mathrm{g\,cm^{-3}}\))&
0.81 & 0.84 & 0.88 &
0.78$^{\mathrm{q}}$ &
0.78$^{\mathrm{q}}$, 0.78$^{\mathrm{t}}$, 0.78$^{\mathrm{u}}$ \\

\hline
\multicolumn{6}{c}{\textbf{NaH}}\\
\hline
$k$ (\(\mathrm{W\,m^{-1}\,K^{-1}}\))&
5.16 & 5.23 & 7.30 &
8.68$^{\mathrm{n}}$, 15.41$^{\mathrm{v}}$ &
5.00$^{\mathrm{w}}$ \\
$c_p$ (\(\mathrm{kJ\,kg^{-1}\,K^{-1}}\))&
1.51 & 1.52 & 1.46 &
-- &
1.52$^{\mathrm{i}}$ \\
$\bar{\gamma}$ &
1.68 & 1.77 & 1.65 &
1.76$^{\mathrm{r}}$, 1.80 (CN)$^{\mathrm{x}}$, 2.20 (QH)$^{\mathrm{x}}$ &
-- \\
$\rho$ (\(\mathrm{g\,cm^{-3}}\))&
1.41 & 1.44 & 1.54 &
1.37$^{\mathrm{q}}$ &
1.39$^{\mathrm{i}}$, 1.40$^{\mathrm{q}}$, 1.37$^{\mathrm{y}}$ \\

\hline
\multicolumn{6}{c}{\textbf{KH}}\\
\hline
$k$ (\(\mathrm{W\,m^{-1}\,K^{-1}}\))&
4.23 & 3.31 & 10.66 &
-- &
-- \\
$c_p$ (\(\mathrm{kJ\,kg^{-1}\,K^{-1}}\))&
0.99 & 0.98 & 0.94 &
-- &
-- \\
$\bar{\gamma}$ &
1.59 & 1.55 & 1.28 &
2.00 (CN)$^{\mathrm{x}}$, 2.20 (QH)$^{\mathrm{x}}$ &
-- \\
$\rho$ (\(\mathrm{g\,cm^{-3}}\))&
1.44 & 1.51 & 1.63 &
1.43$^{\mathrm{q}}$ &
1.43$^{\mathrm{i}}$, 1.43$^{\mathrm{q}}$ \\
\hline
\end{tabularx}

\begin{minipage}{\textwidth}
\footnotesize
$^{\mathrm{a}}$Ref.~\cite{morelli2006high}.
$^{\mathrm{b}}$Ref.~\cite{toher2014high}.
$^{\mathrm{c}}$Ref.~\cite{smirnov1845thermal}.
$^{\mathrm{d}}$Ref.~\cite{touloukian1972thermophysical}.
$^{\mathrm{e}}$Ref.~\cite{kachhava1968semiempirical}.
$^{\mathrm{f}}$Ref.~\cite{spangenberg1957elastischen}.
$^{\mathrm{g}}$Ref.~\cite{lewis1967elastic}.
$^{\mathrm{h}}$Ref.~\cite{miller1964pressure}.
$^{\mathrm{i}}$Ref.~\cite{haynes2016crc}.
$^{\mathrm{j}}$Ref.~\cite{lindsay2016isotope}.
$^{\mathrm{k}}$Ref.~\cite{liang2018lattice}.
$^{\mathrm{l}}$Ref.~\cite{petrov1976behaviour}.
$^{\mathrm{m}}$Ref.~\cite{sirdeshmukh2001alkali}.
$^{\mathrm{n}}$Ref.~\cite{yang2017thermal}.
$^{\mathrm{o}}$Ref.~\cite{vetrano1957batelle}.
$^{\mathrm{p}}$Ref.~\cite{yates1974specific}.
$^{\mathrm{q}}$Ref.~\cite{mueller2013metal}.
$^{\mathrm{r}}$Ref.~\cite{islam1994crystal}.
$^{\mathrm{s}}$Ref.~\cite{gerlich1974pressure}.
$^{\mathrm{t}}$Ref.~\cite{pretzel1960properties}.
$^{\mathrm{u}}$Ref.~\cite{messer1960survey}.
$^{\mathrm{v}}$Ref.~\cite{zhao2017high}.
$^{\mathrm{w}}$Ref.~\cite{bird2020thermal}.
$^{\mathrm{x}}$Ref.~\cite{martins1990equations}.
$^{\mathrm{y}}$Ref.~\cite{shull1948neutron}.
$k^\ast$ at $T=327$\,K.
\end{minipage}

\end{table*}

\begin{figure}[htbp]
  \centering  \includegraphics[width=\linewidth,height=.35\textheight,keepaspectratio]{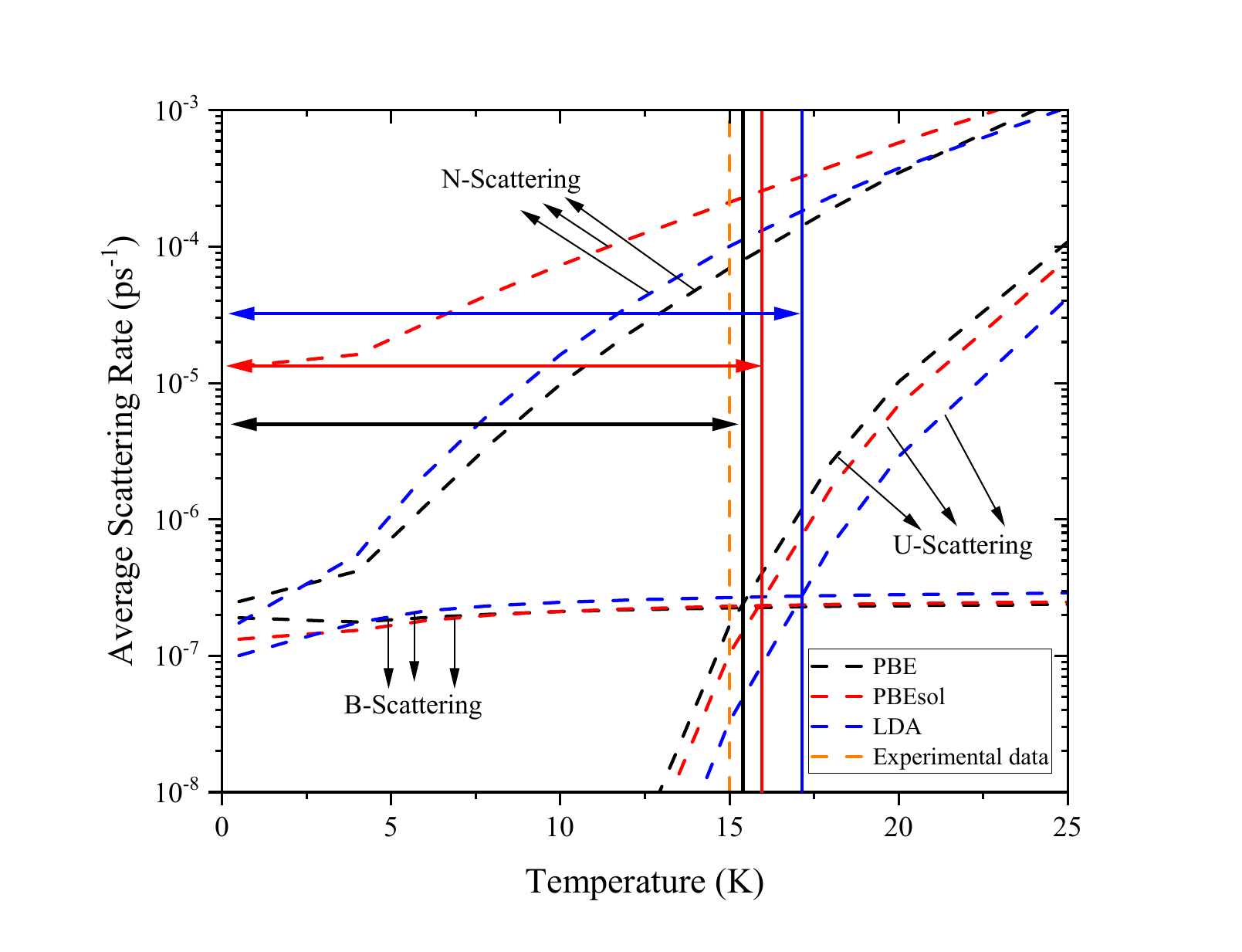}
  \caption{
 Phonon hydrodynamics windows for NaF at $L$ = 8.3 mm. Calculations were done with PBE (black dashed lines), PBEsol (red dashed lines), and LDA (blue dashed lines). Double arrow lines show the hydrodynamics range for each functional. The experimental value (Ref.~\cite{jackson1970second}) is shown as a dashed vertical orange line.}
  \label{fig:NaF_window}
\end{figure}

An example of the application of Guyer’s criterion for $L$ = 8.3 mm is shown in figure \ref{fig:NaF_window}. The hydrodynamic window is located up to 15.5 K for PBE, up to 16 K for PBEsol, and up to 17.2 K for LDA, which agrees with the range of temperatures for the experimental observation of second sound in NaF \cite{jackson1970second}. These ranges indicate the temperature intervals within which each functional predicts phonon hydrodynamics. We characterize phonon hydrodynamics windows at characteristic lengths of $L$ = 10 mm, $L$ = 10 \( \mu\mathrm{m} \),  and $L$ = 10 nm for NaF, LiF, LiH, and NaH (see Appendix E of
the supplementary material). We also present the Normal and Umklapp scattering rates at $T$ = 20 K and $T$ = 300 K (see Appendix D of the supplementary material).

\begin{figure}[p]
  \centering
  \includegraphics[page=1,
                   width=\textwidth,
                   height=0.4\textheight,
                   keepaspectratio]{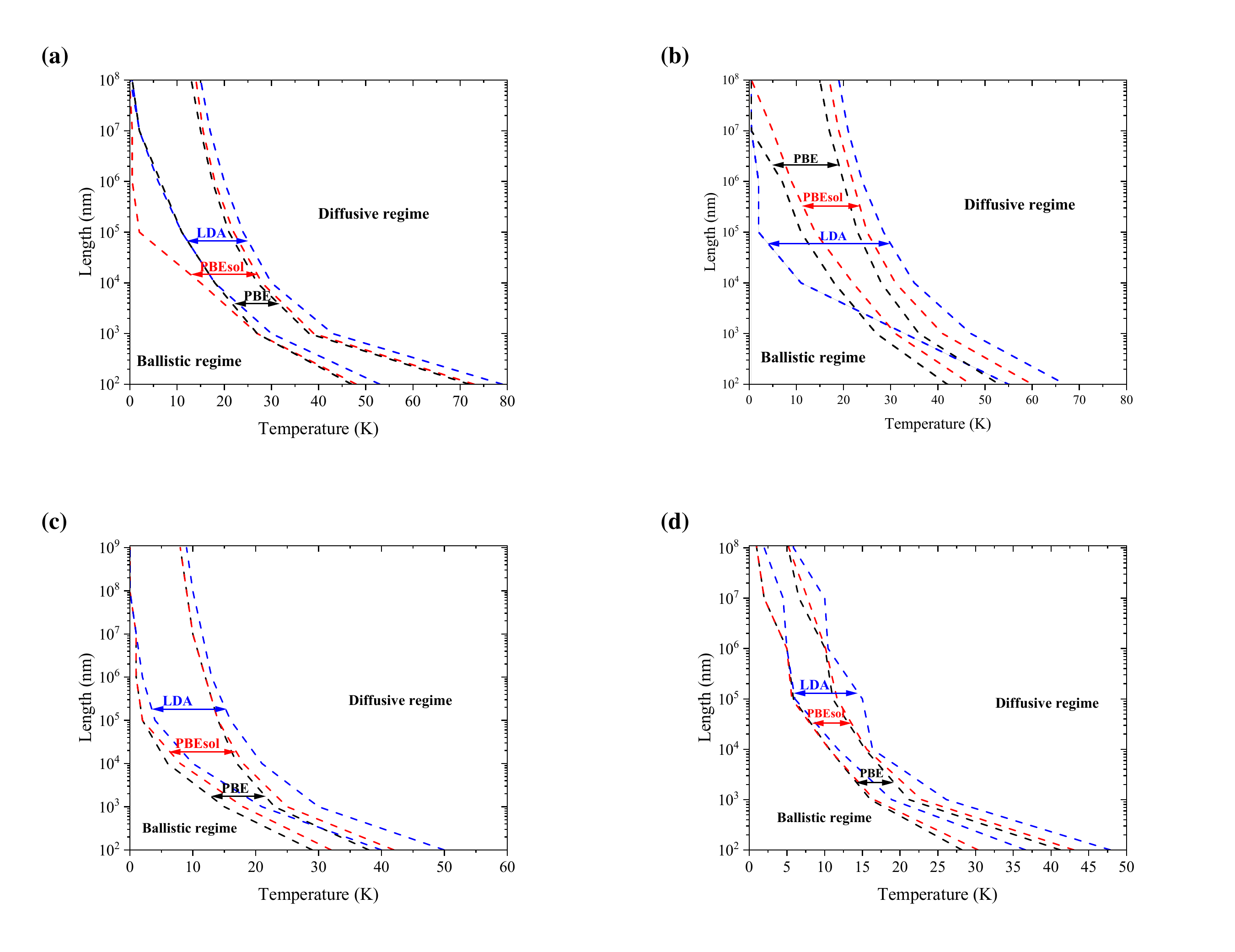}

  \includegraphics[page=2,
                   width=\textwidth,
                   height=0.4\textheight,
                   keepaspectratio]{hydrodynamics-all}
  
  \caption{
  Thermal transport regimes (phonon hydrodynamics regime, ballistic regime, and diffusive regime) for: (a) NaF. (b) LiF. (c) KF. (d) NaCl. (e) KCl. (f) LiH. (g) NaH. (h) KH. Double arrow lines show the window of phonon hydrodynamics for each functional. PBE (black dashed lines), PBEsol (red dashed lines), and LDA (blue dashed lines).}
  \label{fig:hydrodynamics}
\end{figure}

By applying Guyer’s criterion at different temperatures and characteristic lengths, the ‘phase space’ of phonon transport regimes can be determined. figure \ref{fig:hydrodynamics} shows the three phonon transport regimes (ballistic, hydrodynamic, and diffusive) for the eight fluorides, chlorides, and hydrides, over temperatures up to \(80\,\mathrm{K}\) and characteristic lengths from \(10^2\) to \(10^8\,\mathrm{nm}\). To the best of our knowledge, these are the first reported predictions of phonon hydrodynamics in LiH, NaH, KH, KF, KCl and NaCl. Additionally, we solve the linearized BTE with the full scattering matrix operator for NaF, NaH, LiF, KF and KH to obtain predictions for the observation of second sound (see supplementary material Appendix H).  LDA generally yields the widest phonon-hydrodynamic window because, relative to PBE/PBEsol, it shifts the phonon properties in a way that makes Normal scattering dominate over resistive scattering across a broader temperature interval. LiF exhibits the broadest phonon hydrodynamic window, whereas KCl and NaH show the narrowest. 

\newpage
\clearpage

Finally, we briefly comment on the impact of isotopes on the phonon hydrodynamics windows. Resistive scattering rate (R-scattering) can be used instead of Umklapp scattering rate (U-scattering) to predict phonon hydrodynamics by applying Eq. \ref{eq:resistive-criterion}. We report the impact of isotopes on the phonon hydrodynamics windows of NaF, LiF, LiH, and NaH (see figure \ref{fig:isotope_windows_lif_lih} and figure S15 in the supplementary material. For fluorides and chlorides, we notice the hydrodynamics window range at $L$ = 1 mm of LiF narrows to 9 K (shaded red region), compared with U-scattering, which extends up to 24.5 K (shaded transparent red region). This is consistent with the analysis from Ackerman and Guyer ~\cite{ackerman1968temperature}, who argue that high crystal purity is required to observe second sound in LiF. On the other hand, the curves of U-scattering and R-scattering in NaF remain close to each other, and consequently, the hydrodynamics windows are nearly the same. This small impact of isotopes may explain the success of the early experimental work on NaF which confirmed the existence of second sound~\cite{jackson1970second,mcnelly1970heat}. Likewise, NaH remains unaffected by isotope scattering. On the other hand, the hydrodynamic windows disappear at $L$ = 1 mm for LiH when isotope scattering is included.


\begin{figure}[htbp]
  \centering  \includegraphics[width=\linewidth,height=.35\textheight,keepaspectratio]{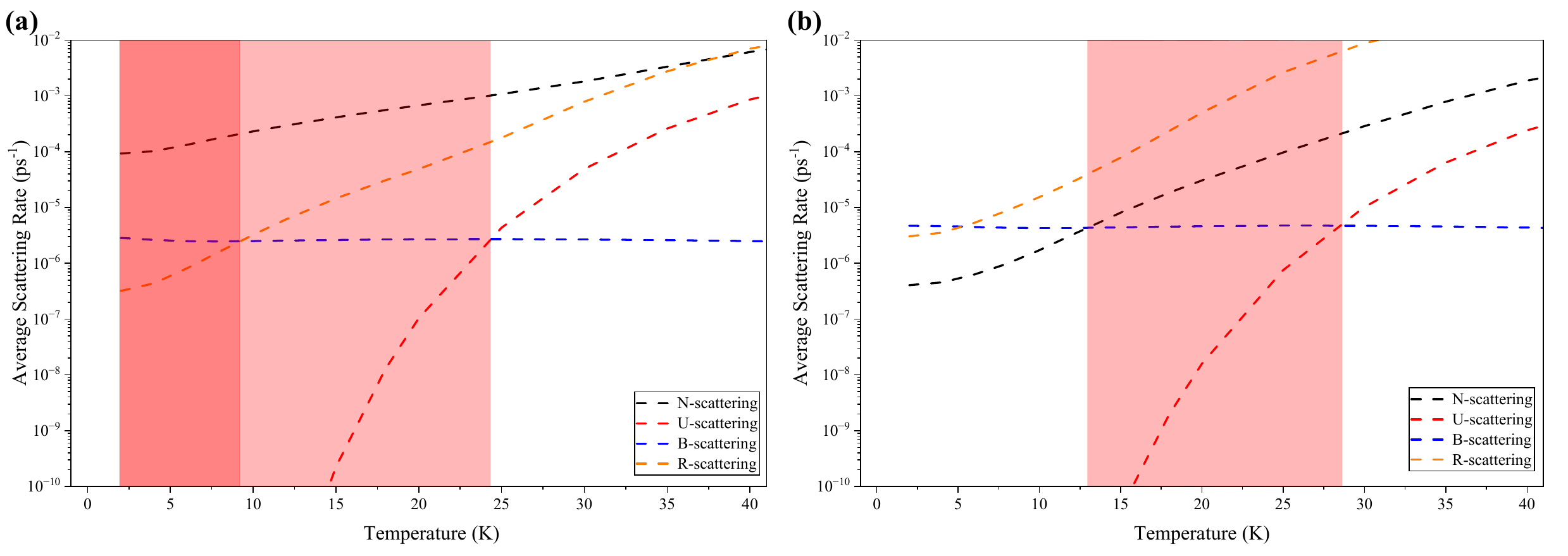}
  \caption{Phonon hydrodynamics windows for pure materials (shaded transparent red region) and with impurity (shaded red region) at $L$ = 1 mm for: (a) LiF using LDA. (b) LiH using PBE. The scattering rates are denoted by colors: N-scattering (black dashed lines), U-scattering (red dashed lines), B-scattering (blue dashed lines), and R-scattering (orange dashed lines)}
  \label{fig:isotope_windows_lif_lih}
\end{figure}

\newpage
\clearpage

\section{Discussion}

\begin{figure}[htbp]
  \centering  \includegraphics[width=\linewidth,height=.35\textheight,keepaspectratio]{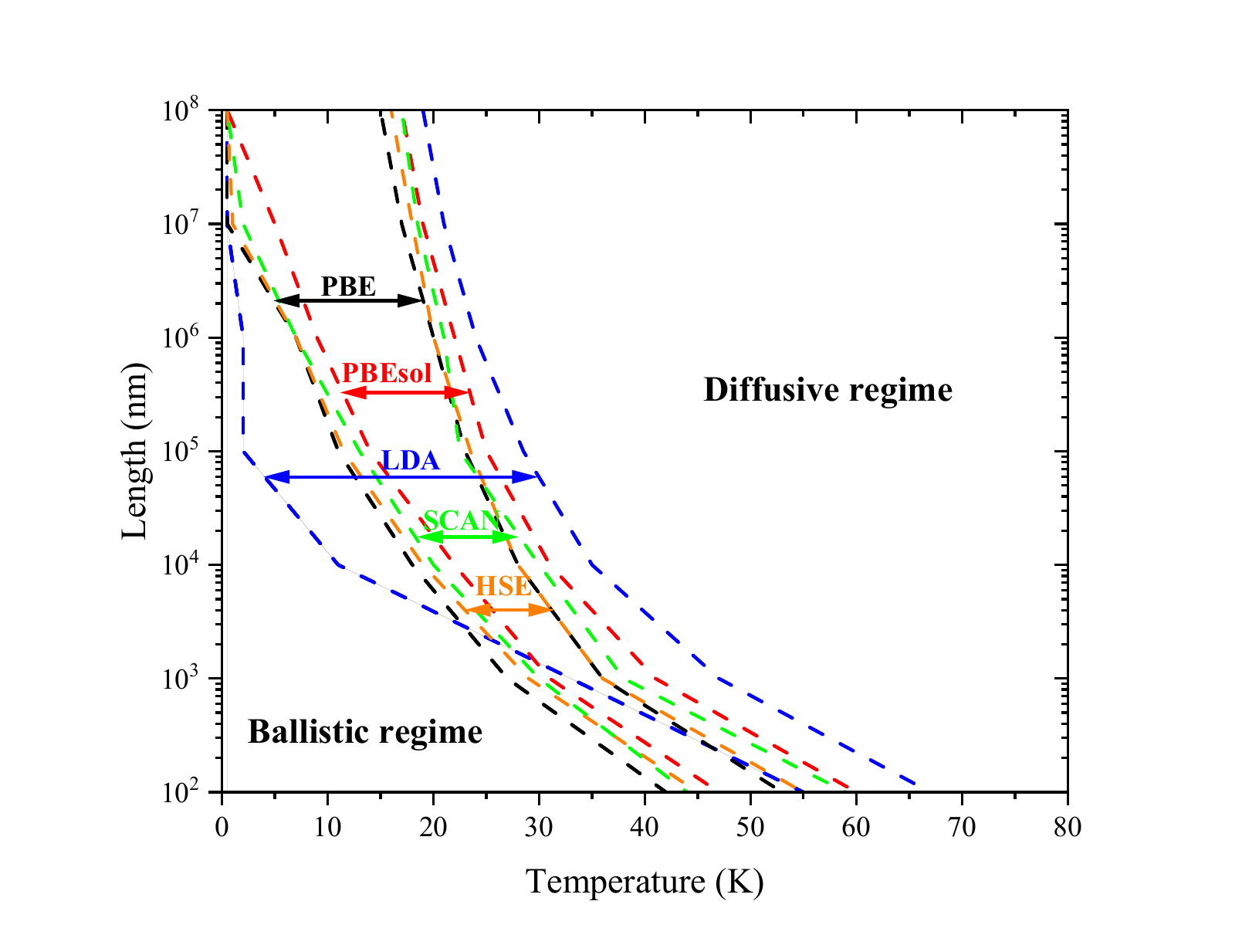}
  \caption{Thermal transport regimes for LiF. Double arrow lines show the window of phonon hydrodynamics for each functional. PBE (black dashed lines), PBEsol (red dashed lines), LDA (blue dashed lines), SCAN (green dashed lines), and HSE (orange dashed lines).}
  \label{fig:comp.transport}
\end{figure}

To reconcile the observation that the calculations of electronic band gap deviate significantly from experimental values, while better agreement is obtained for phonon properties, we examine a selection of modern XC functionals. For instance, Becke and Johnson (BJ) exchange potential was proposed to precisely regenerate the exact atomic exchange optimized effective potential (OEP) in terms of shape, resulting in accurately calculated electronic band gaps \cite{tran2007band}. To assess band-gap accuracy using this functional, we evaluated LiF, NaCl, and LiH, obtaining 11.26, 7.81, and 4.80 eV, respectively, in agreement with measured values \cite{roessler1967electronic, piacentini1976thermoreflectance, miyata1968optical, himpsel1978angle, kondo1988effect, plekhanov1998wannier}. However, this functional is less accurate in terms of structural calculations in comparison to PBEsol. The calculated lattice constants are 4.08 \AA{} for LiF, 5.69 \AA{} for NaCl, and 3.97 \AA{} for LiH. A more accurate choice for structural calculations is the SCAN functional \cite{sun2015scan}. The calculated lattice constants using the SCAN functional are 3.97 \AA{} for LiF, 5.59 \AA{} for NaCl, and 4.00 \AA{} for LiH. Nevertheless, the calculated band gap using SCAN is less accurate than the BJ functional (LiF =  10.27 eV, NaCl = 6.10 eV, LiH = 3.58 eV). The HSE (Heyd-Scuseria-Ernzerhof) hybrid functional \cite{heyd2003hybrid} is another functional that has been shown to improve electronic properties such as band gaps. We tested PBE with HSE in LiF and found that the band gap increased to 11.17 eV with an error of $\sim 18\%$. Since the HSE functional is not presently implemented within the ph.x package of Quantum ESPRESSO \cite{giannozzi2009quantum}, Phonopy \cite{togo2023implementation, togo2023first} and Phono3py \cite{togo2023implementation, togo2023first} were employed to calculate the second and third interatomic force constants, respectively. To ensure a fair comparison among exchange–correlation functionals, we assess the predicted thermal-transport regimes (see figure \ref{fig:comp.transport}). Both HSE and SCAN trends fall between PBE and PBEsol. 

One possible reason phonon calculations remain weakly, but non-negligibly impacted, by the choice of functional and thereby ensure the reasonable accuracy of thermal calculations in contrast to electronic calculations, is the distinct impact of absolute total energies on electronic calculations and the influence of relative energy differences on phonon calculations \cite{carbogno2022numerical}. Absolute total energies in electronic calculations suffer from inaccuracies due to the approximations made within the XC functionals. On the other hand, phonon calculations (i.e., force constants) depend on energy differences, which can be computed more reliably since any consistent error or systematic offset in energy calculations may cancel out, resulting in an accurate description of thermal properties. In other words, phonon properties depend on the shape of the interatomic potential energy surface and not its absolute value. Even so, the influence of exchange-correlation functionals on phonon properties is evident as demonstrated in the variability of the phonon hydrodynamics windows for all the materials studied in this work.

It is insightful to consider the steps required to produce reliable first-principles calculations in the absence of experimental data. As a starting point, one can begin with materials with similar atomic compositions for which functionals have been developed and validated. In the absence of a well-established pseudopotential, one can obtain a preliminary pseudopotential using a pseudopotential generation code, such as those based on the projector-augmented wave (PAW) method. This preliminary pseudopotential can be refined through iterative calculations, comparing the results with available experimental and theoretical data in related systems, even if indirect. Convergence of DFT input parameters, such as k-point sampling and cutoff energy, is required to remove additional sources of uncertainty. Additionally, beyond the selection of pseudopotential and exchange-correlation functional, there is ongoing work to benchmark the large number of available DFT codes. For instance, Bosoni \textit{et al} \cite{bosoni2024verify} proposed a protocol to enhance the reliability and precision of DFT methods and codes, providing a substantial reference dataset to foster improvements and ensure reproducibility in computational studies. The protocol leverages AiiDA workflows to streamline and ensure reproducibility in verifying DFT computations using a dataset based on all-electron methods spanning the entire periodic table \cite{bosoni2024verify}. This effort can be interpreted as a way to measure numerical error (i.e., choice of algorithms and not the choice of XC functional) in DFT codes, which can be combined with error due to the functional approximations via error propagation approaches \cite{parks2020uncertainty}.

\section{Conclusions}

Using DFT, we calculated the electrical, mechanical, and thermal properties of NaF, LiF, KF, KCl, NaCl, LiH, NaH, and KH  using PBE, PBEsol, and LDA functionals. Phonon scattering rates were used to predict phonon hydrodynamics windows based on Guyer's condition and solutions to the linearized BTE. The impact of isotopes on lattice thermal properties and phonon hydrodynamics windows was characterized. We find that GGA functionals, when compared to the LDA functionals, do not necessarily lead to more accurate predictions of the thermal conductivity. Overall, our calculations demonstrate good agreement with, when available, previous experimental and previous theoretical results. We find non-negligible differences in the prediction of the phonon hydrodynamics window due to the choice of functional. Building on recent first-principles studies that explicitly include four-phonon scattering in BTE calculations \cite{han2022fourphonon, tang2023strong, tang2024effects}, incorporating four-phonon processes in future work would further improve quantitative accuracy by refining the resulting average phonon scattering rates.

\section*{Data availability statement}

All data that support the findings of this study are included within the article (and any supplementary files).

\section*{Acknowledgements}
This work was supported by the NSERC Discovery Grants Program under Grant No. RGPIN-2021-02957 and FRQNT Nouveau Chercheur No. 341503.

\section*{Conflict of interest}

The authors declare no financial competing interest.

\section*{Author contributions}

Jamal Abou Haibeh ~\orcidlink{0009-0003-7167-4731} \href{https://orcid.org/0009-0003-7167-4731}{0009-0003-7167-4731}

Conceptualization (Equal); Data curation (Lead); Formal analysis (Lead); Investigation (Equal); Methodology (Equal); Validation (Equal); Visualization (Lead); Writing – original draft (Lead); Writing – review \& editing (Equal).

Samuel Huberman ~\orcidlink{0000-0003-0865-8096} 
\href{https://orcid.org/0000-0003-0865-8096}{0000-0003-0865-8096}

Conceptualization (Equal); Data curation (Supporting); Formal analysis (Supporting); Funding acquisition (Lead); Investigation (Equal); Methodology (Equal); Project administration (Lead); Resources (Lead); Software (Lead); Supervision (Lead); Validation (Equal); Visualization (Supporting); Writing – original draft (Supporting); Writing – review \& editing (Equal).

\bibliographystyle{iopart-num}

\bibliography{references}

\end{document}


\title{Supplementary Material for: Impact of Exchange-Correlation Functionals on Predictions of Phonon Hydrodynamics: A Study of Fluorides, Chlorides, and Hydrides}

\author{Jamal Abou Haibeh$^1$~\orcidlink{0009-0003-7167-4731} and Samuel Huberman$^{1,2,*}$~\orcidlink{0000-0003-0865-8096}\thanks{Corresponding author: \href{samuel.huberman@mcgill.ca}{samuel.huberman@mcgill.ca}}}

\affil{$^1$Department of Chemical Engineering, McGill University, Montreal, Quebec H3A 0C5, Canada}

\affil{$^2$Department of Physics, McGill University, Montreal, Quebec H3A 0C5, Canada}

\date{}

\maketitle

\phantomsection
\label{supp:doc}

\setcounter{section}{0}
\renewcommand{\thesection}{S\Roman{section}}

\setcounter{figure}{0}
\renewcommand{\thefigure}{S\arabic{figure}}

\setcounter{table}{0}
\renewcommand{\thetable}{S\arabic{table}}

\setcounter{equation}{0}
\renewcommand{\theequation}{S\arabic{equation}}

\appendix

\section{DFT calculations}\label{sec:dft}

The DFT calculations were done using PBE, PBEsol, and LDA exchange-correlation functionals. All the studied structures of fluorides, chlorides, and hydrides, including NaF, LiF, KF, NaCl, KCl, LiH, NaH, and KH, adopt the face-centered cubic (FCC) crystal structure. For first-principles calculations, a 14 × 14 × 14 k-point mesh of Monkhorst-Pack with a 100 Ry kinetic energy cutoff and a convergence threshold of 1.0E-16 Ry was chosen. Quantum ESPRESSO package \cite{giannozzi2009quantum} was used to obtain second-order interatomic force constants, whereas thirdorder.py \cite{li2014shengbte} was used to obtain third-order interatomic force constants. BTE was solved using the ShengBTE package \cite{li2014shengbte}. The phonon Boltzmann transport equation was solved using the iterative approach to obtain the thermal transport properties. Thermal conductivity was converged on a 40 × 40 × 40 q-mesh with a Gaussian smearing parameter of 1.0 for the Kronecker delta approximation. All data, including input files and interatomic force constants, is freely available on our public repository \footnote{\url{https://github.com/netlabmcgill/dft_calculations}}.

\clearpage
\Needspace{0.95\textheight}
\section{Electronic band structures}\label{sec:band}

\begin{figure}[H]
  \centering

  \includegraphics[page=1,
    width=\linewidth,
    height=0.4\textheight,
    keepaspectratio,
    trim=0 6mm 0 4mm,clip]{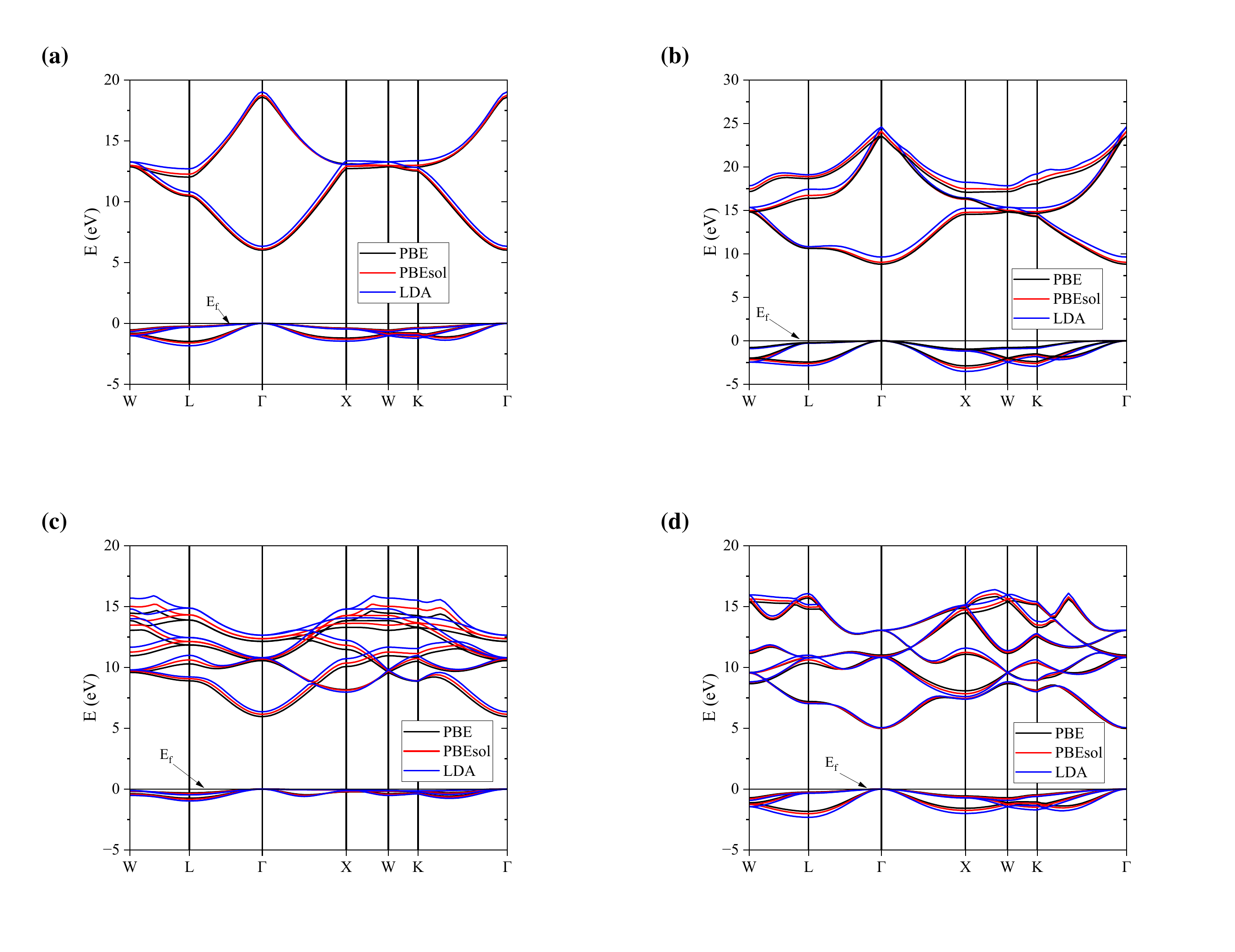}

  \vspace{2mm}

  \includegraphics[page=2,
    width=\linewidth,
    height=0.4\textheight,
    keepaspectratio,
    trim=0 6mm 0 4mm,clip]{electronic_band-all}

  \caption{Electronic band structures of: (a) NaF. (b) LiF. (c) KF. (d) NaCl. (e) KCl. (f) LiH. (g) NaH. (h) KH.}
  \label{fig:band}
\end{figure}

\newpage
\clearpage

\Needspace{0.92\textheight}
\section{Average phonon scattering rates at $T$ = 20 K versus q-point mesh (N$\times$N$\times$N)}\label{sec:conv}

\begin{figure}[H]
  \centering
  \includegraphics[width=\linewidth,height=0.8\textheight,keepaspectratio]{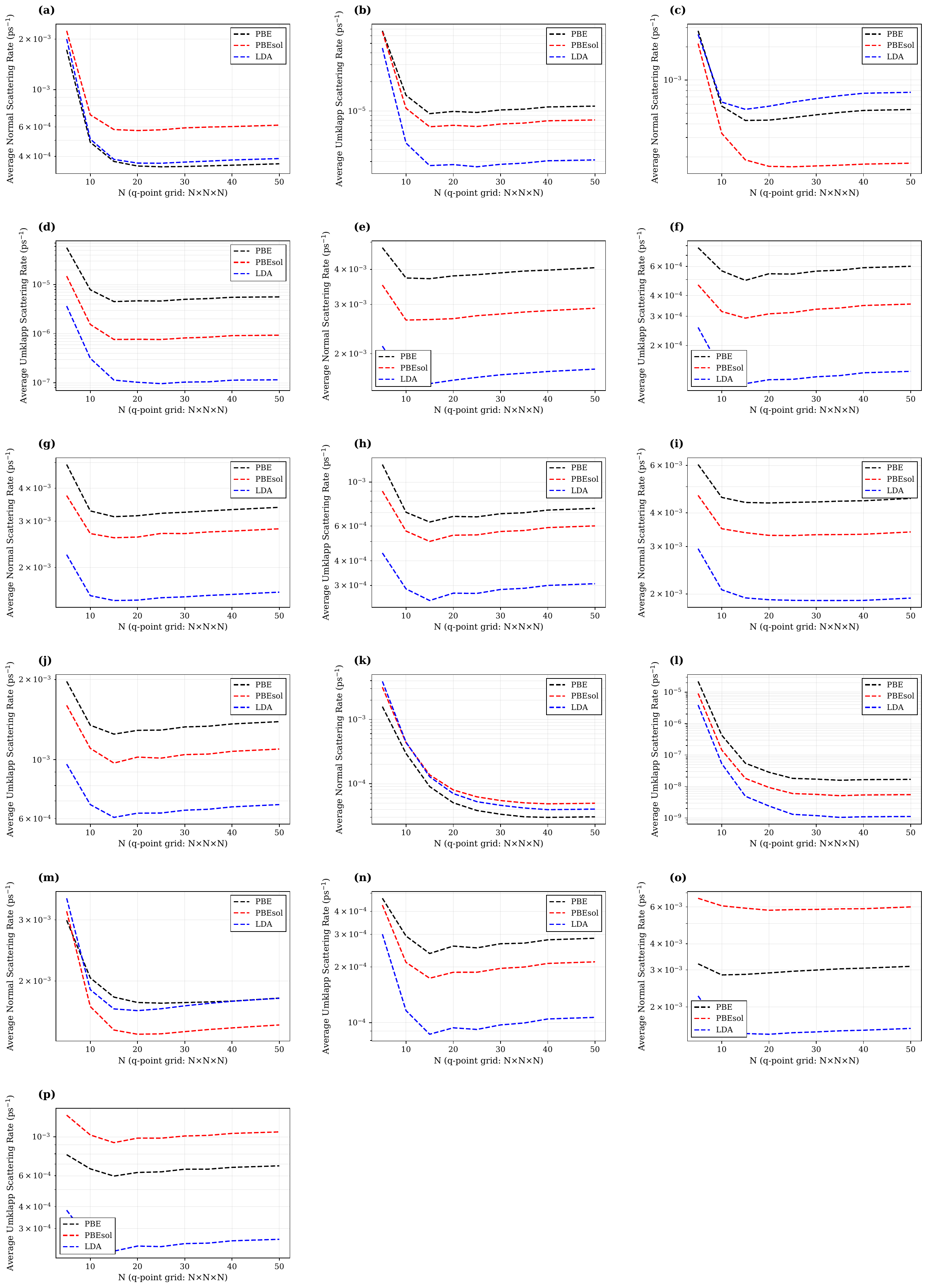}
  \caption{Average phonon scattering rates at $T$ = 20 K as a function of q-point mesh. Calculations were done with PBE (black dashed lines), PBEsol (red dashed lines), and LDA (blue dashed lines). Panels are ordered as: (a) NaF–Normal, (b) NaF–Umklapp, (c) LiF–Normal, (d) LiF–Umklapp, (e) KF–Normal, (f) KF–Umklapp, (g) NaCl–Normal, (h) NaCl–Umklapp, (i) KCl–Normal, (j) KCl–Umklapp, (k) LiH–Normal, (l) LiH–Umklapp, (m) NaH–Normal, (n) NaH–Umklapp, (o) KH–Normal, (p) KH–Umklapp.}
  \label{fig:conv}
\end{figure}

\newpage
\clearpage

\section{Normal and Umklapp scattering rates at $T$ = 20 K and $T$ = 300 K}\label{sec:scattering}

\textbf{(1) NaF:
}
\begin{figure}[htbp]
  \centering  \includegraphics[width=\linewidth,height=.5\textheight,keepaspectratio]{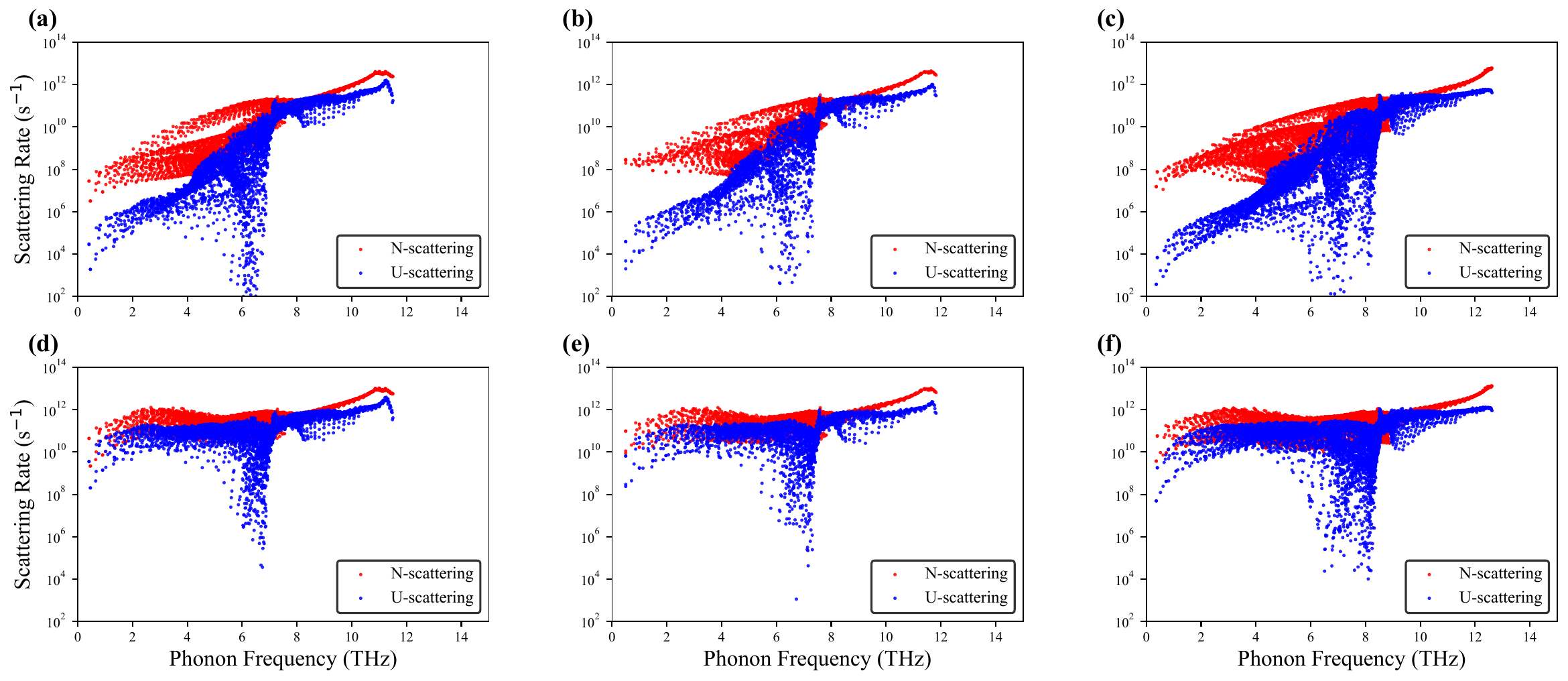}
  \caption{Normal (red circles) and Umklapp (blue circles) scattering rates for NaF at $T = 20,\mathrm{K}$: (a) PBE, (b) PBEsol, and (c) LDA; and at $T = 300,\mathrm{K}$: (d) PBE, (e) PBEsol, and (f) LDA.}
  \label{fig:scattering1}
\end{figure}

\textbf{(2) LiF:
}
\begin{figure}[htbp]
  \centering  \includegraphics[width=\linewidth,height=.5\textheight,keepaspectratio]{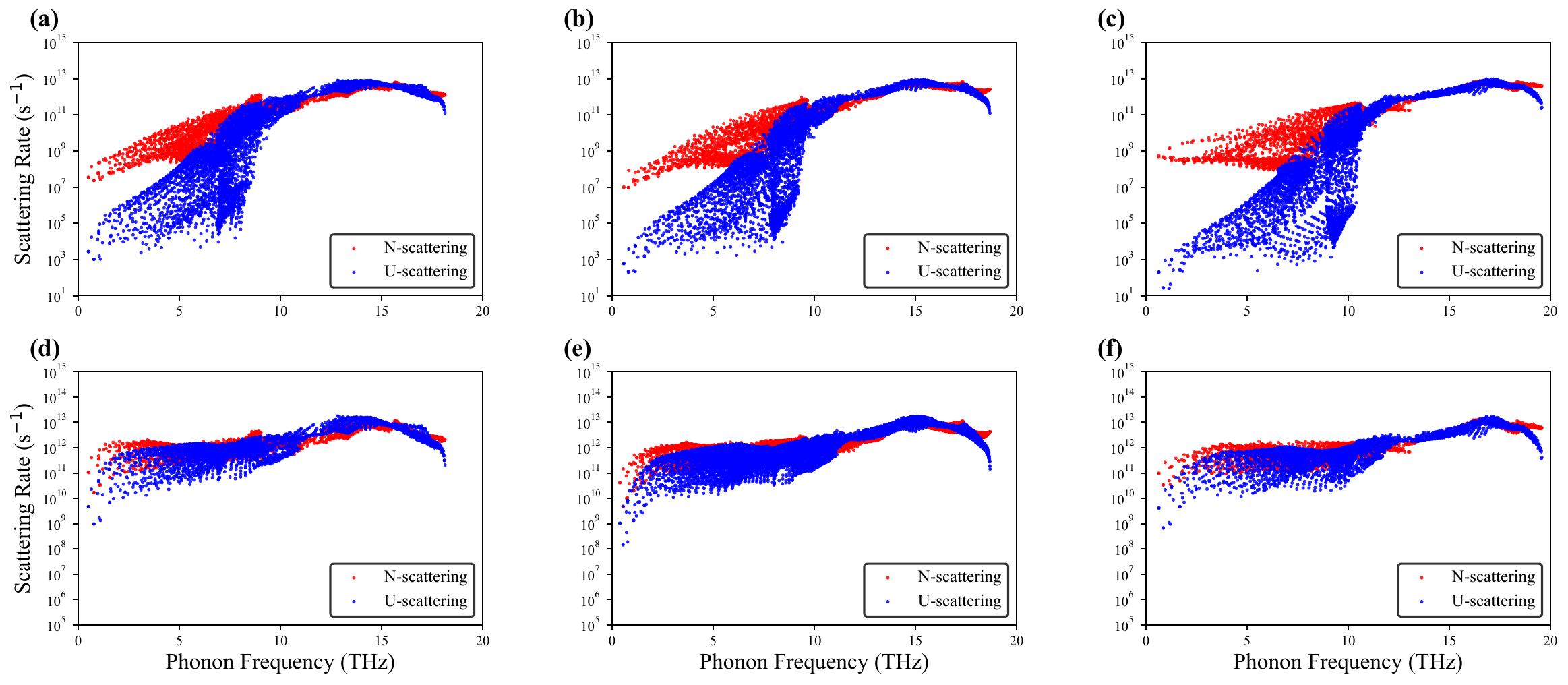}
  \caption{Normal (red circles) and Umklapp (blue circles) scattering rates for LiF at $T = 20,\mathrm{K}$: (a) PBE, (b) PBEsol, and (c) LDA; and at $T = 300,\mathrm{K}$: (d) PBE, (e) PBEsol, and (f) LDA.}
  \label{fig:scattering2}
\end{figure}

\newpage
\clearpage

\textbf{(3) KF:
}
\begin{figure}[htbp]
  \centering  \includegraphics[width=\linewidth,height=.5\textheight,keepaspectratio]{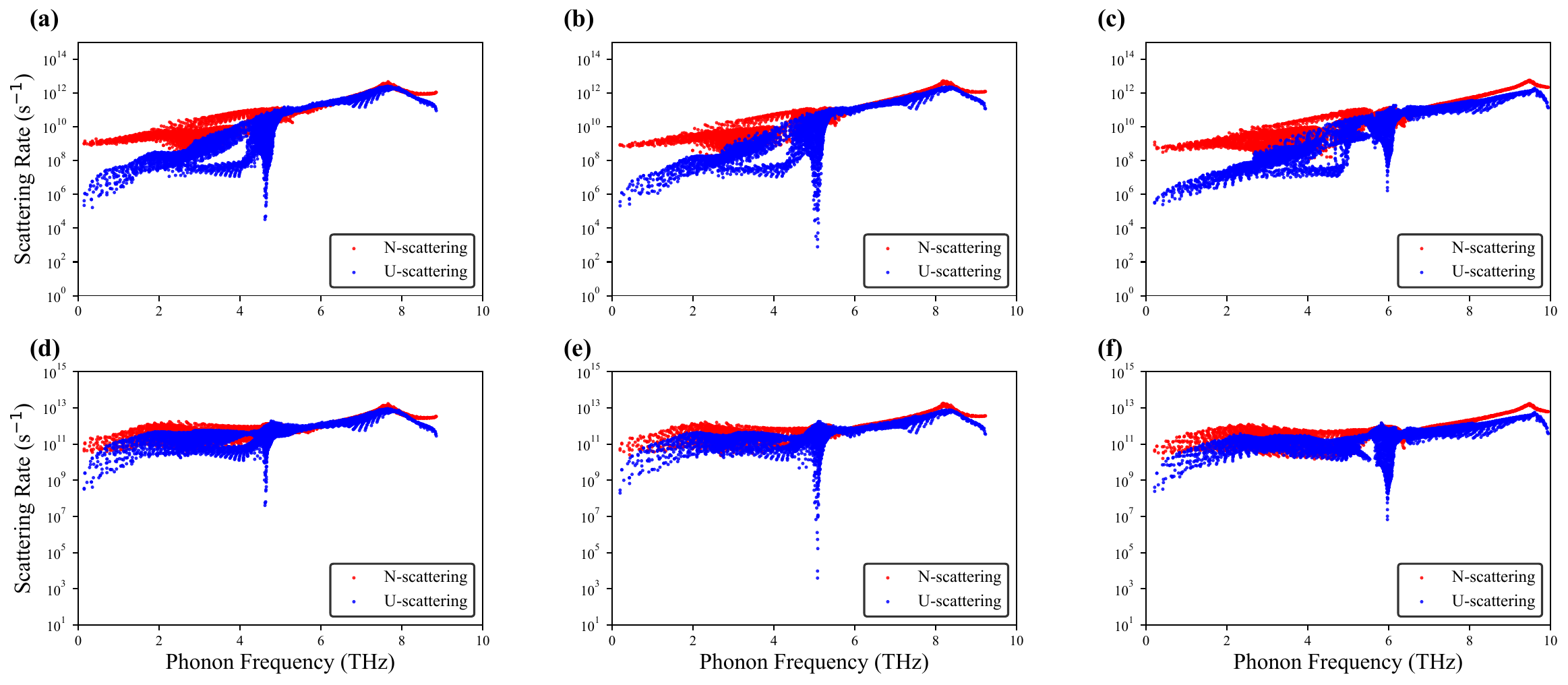}
  \caption{Normal (red circles) and Umklapp (blue circles) scattering rates for KF at $T = 20,\mathrm{K}$: (a) PBE, (b) PBEsol, and (c) LDA; and at $T = 300,\mathrm{K}$: (d) PBE, (e) PBEsol, and (f) LDA.}
  \label{fig:scattering3}
\end{figure}

\textbf{(4) NaCl:
}
\begin{figure}[htbp]
  \centering  \includegraphics[width=\linewidth,height=.5\textheight,keepaspectratio]{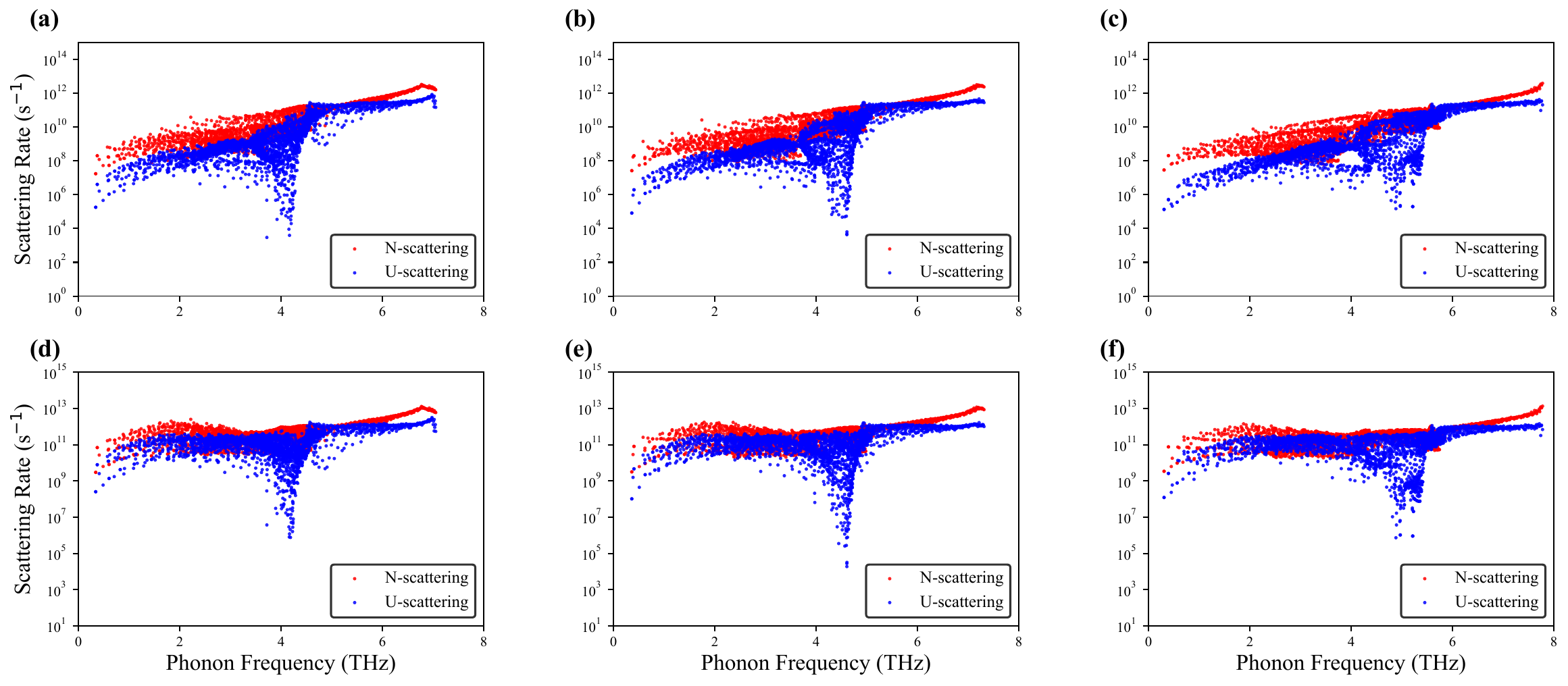}
  \caption{Normal (red circles) and Umklapp (blue circles) scattering rates for NaCl at $T = 20,\mathrm{K}$: (a) PBE, (b) PBEsol, and (c) LDA; and at $T = 300,\mathrm{K}$: (d) PBE, (e) PBEsol, and (f) LDA.}
  \label{fig:scattering4}
\end{figure}

\newpage
\clearpage

\textbf{(5) KCl:
}
\begin{figure}[htbp]
  \centering  \includegraphics[width=\linewidth,height=.5\textheight,keepaspectratio]{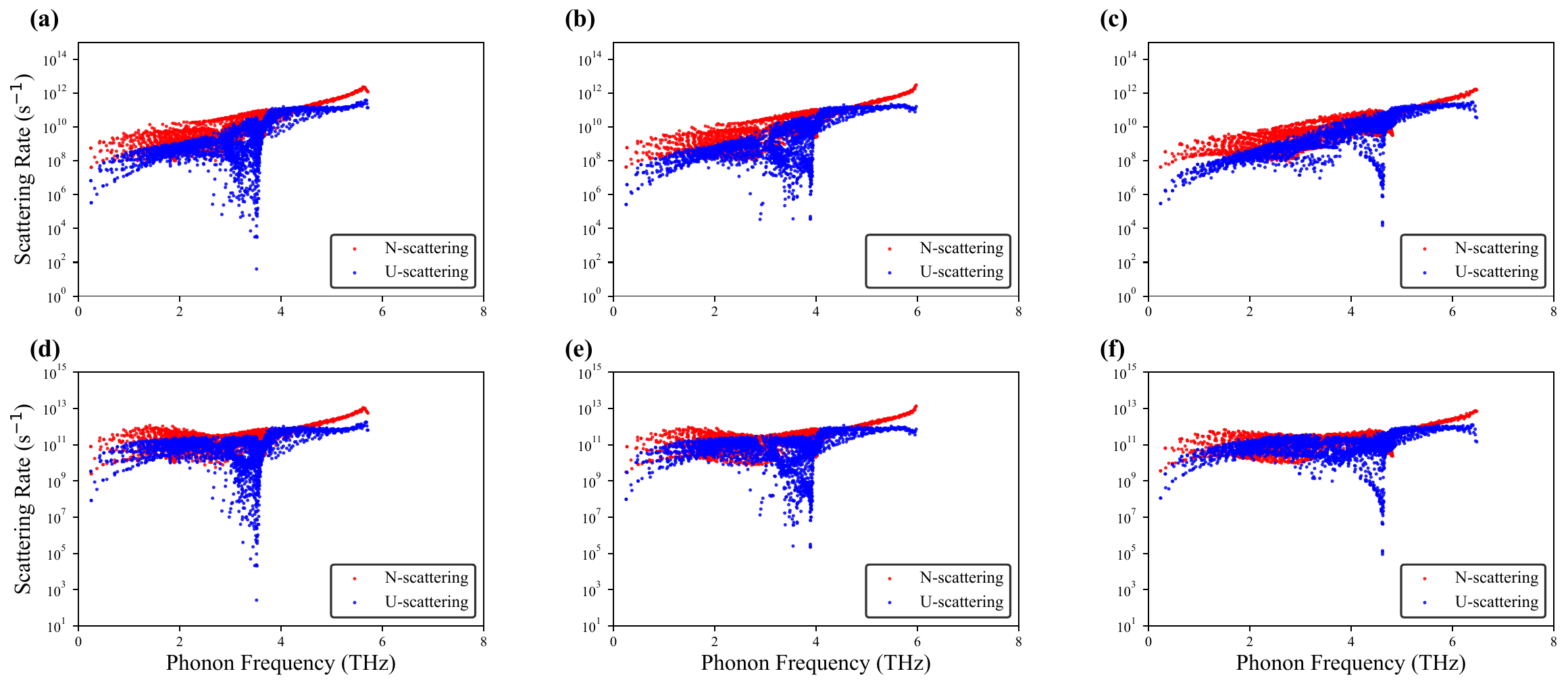}
  \caption{Normal (red circles) and Umklapp (blue circles) scattering rates for KCl at $T = 20,\mathrm{K}$: (a) PBE, (b) PBEsol, and (c) LDA; and at $T = 300,\mathrm{K}$: (d) PBE, (e) PBEsol, and (f) LDA.}
  \label{fig:scattering5}
\end{figure}

\textbf{(6) LiH:
}
\begin{figure}[htbp]
  \centering  \includegraphics[width=\linewidth,height=.5\textheight,keepaspectratio]{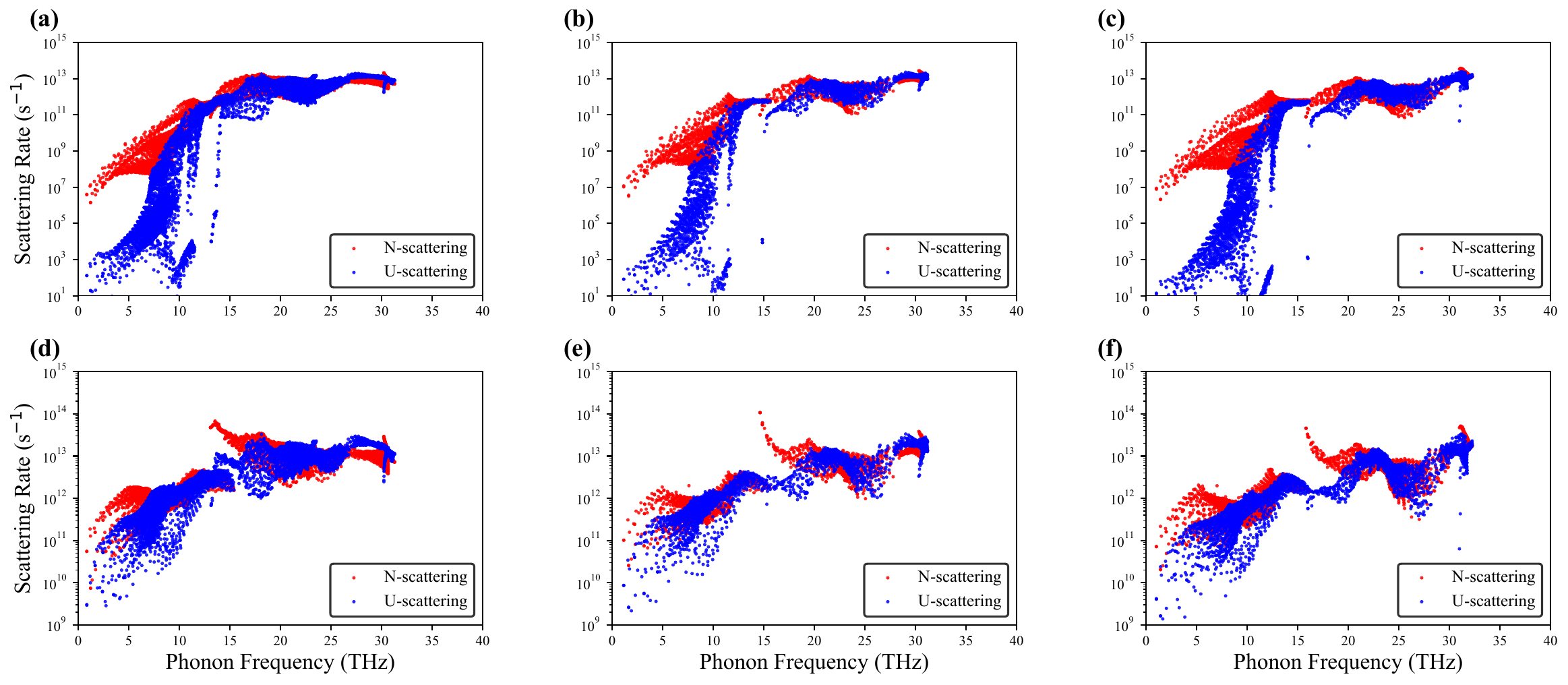}
  \caption{Normal (red circles) and Umklapp (blue circles) scattering rates for LiH at $T = 20,\mathrm{K}$: (a) PBE, (b) PBEsol, and (c) LDA; and at $T = 300,\mathrm{K}$: (d) PBE, (e) PBEsol, and (f) LDA.}
  \label{fig:scattering6}
\end{figure}

\newpage
\clearpage

\textbf{(7) NaH:
}
\begin{figure}[htbp]
  \centering  \includegraphics[width=\linewidth,height=.5\textheight,keepaspectratio]{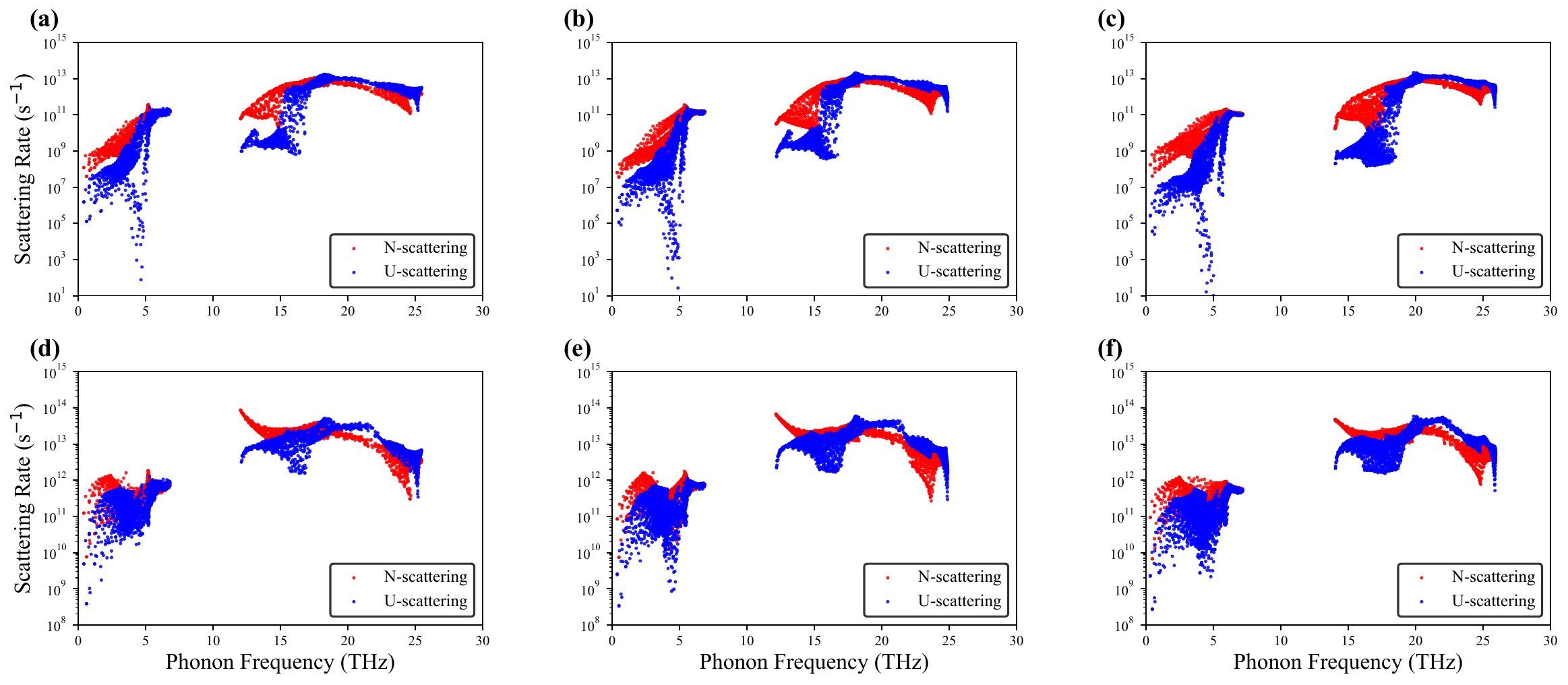}
  \caption{Normal (red circles) and Umklapp (blue circles) scattering rates for NaH at $T = 20,\mathrm{K}$: (a) PBE, (b) PBEsol, and (c) LDA; and at $T = 300,\mathrm{K}$: (d) PBE, (e) PBEsol, and (f) LDA.}
  \label{fig:scattering7}
\end{figure}

\textbf{(8) KH:
}
\begin{figure}[htbp]
  \centering  \includegraphics[width=\linewidth,height=.5\textheight,keepaspectratio]{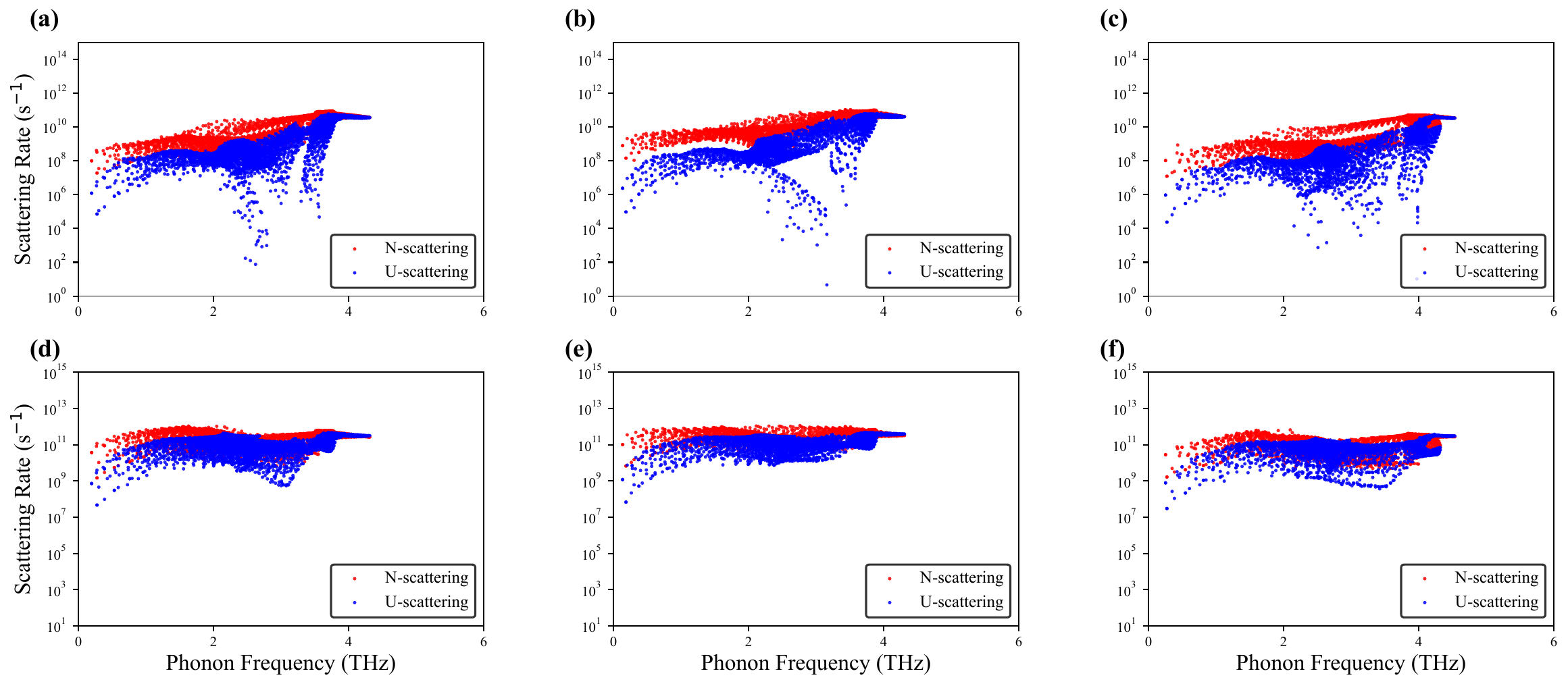}
  \caption{Normal (red circles) and Umklapp (blue circles) scattering rates for KH at $T = 20,\mathrm{K}$: (a) PBE, (b) PBEsol, and (c) LDA; and at $T = 300,\mathrm{K}$: (d) PBE, (e) PBEsol, and (f) LDA.}
  \label{fig:scattering8}
\end{figure}


\newpage
\clearpage

\section{Phonon hydrodynamics windows for pure crystals }\label{sec:phonon_windows}

\textbf{(1) NaF:
}

\begin{figure}[htbp]
  \centering  \includegraphics[width=\linewidth,height=.5\textheight,keepaspectratio]{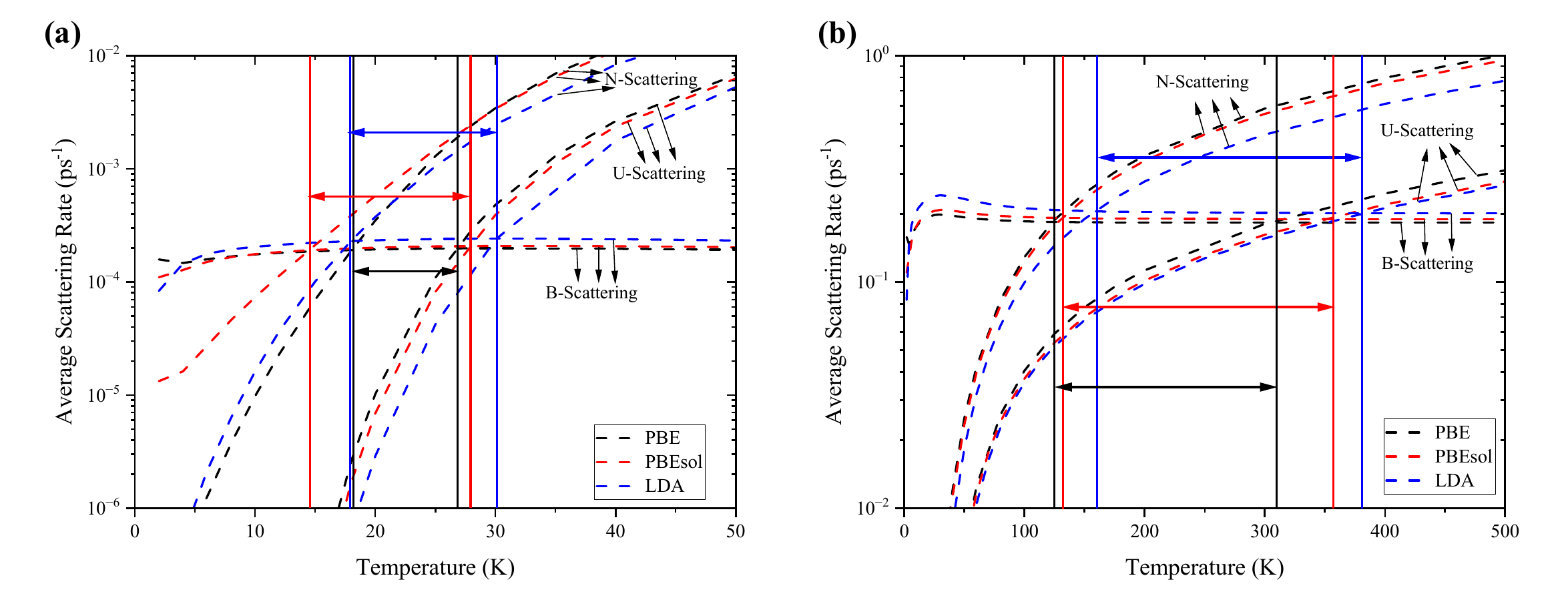}
  \caption{Phonon hydrodynamics windows for NaF. Calculations were done with PBE (black dashed lines), PBEsol (red dashed lines), and LDA (blue dashed lines). Double arrow lines show the window of phonon hydrodynamics for each functional. (a) At $L$ = 10\( \mu\mathrm{m} \). (b) At $L$ = 10 nm.}
  \label{fig:window_naf}
\end{figure}

\newpage
\clearpage

\textbf{(2) LiF:
}

\begin{figure}[htbp]
  \centering  \includegraphics[width=\linewidth,height=.65\textheight,keepaspectratio]{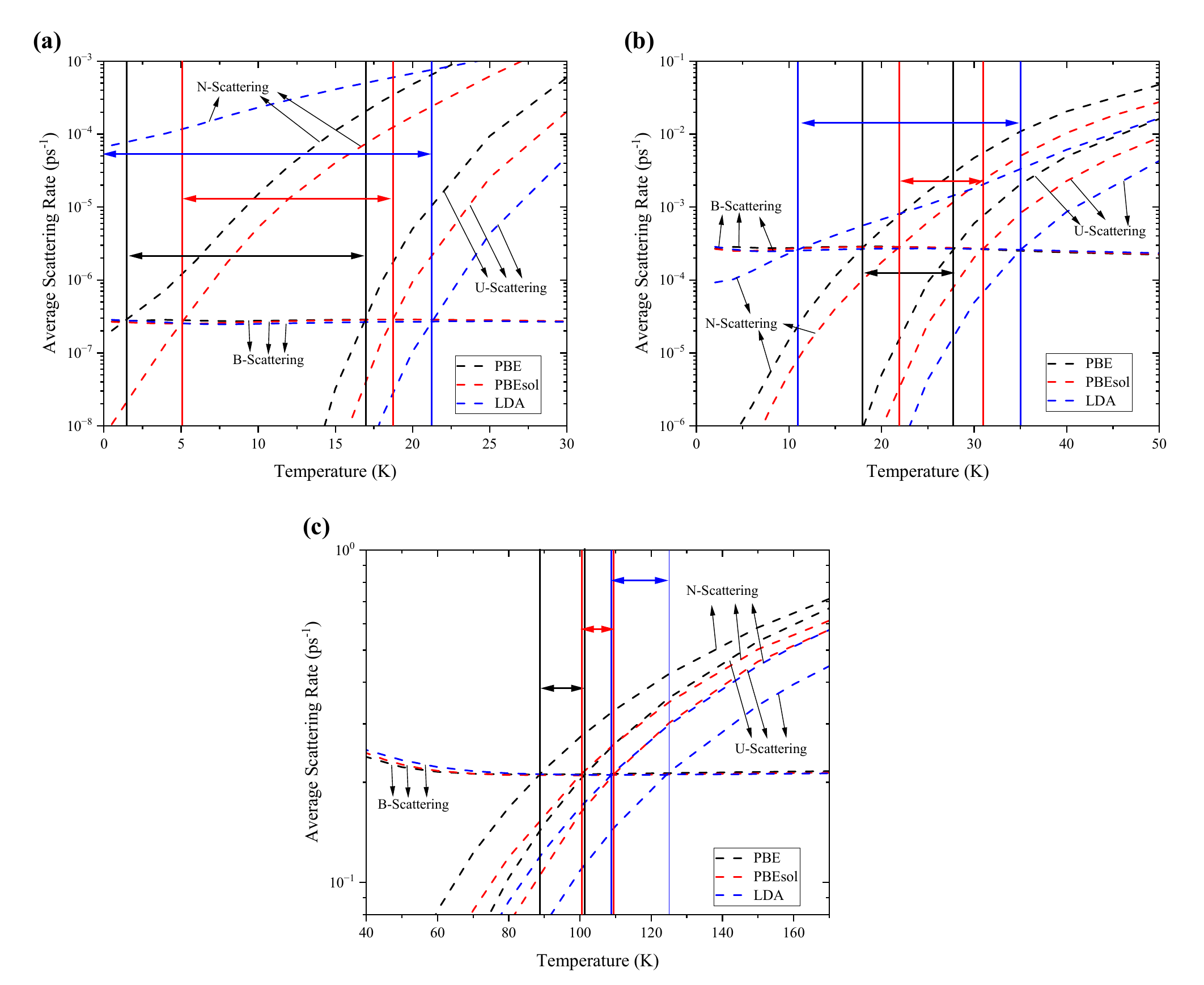}
  \caption{Phonon hydrodynamics windows for LiF. Calculations were done with PBE (black dashed lines), PBEsol (red dashed lines), and LDA (blue dashed lines). Double arrow lines show the window of phonon hydrodynamics for each functional. (a) At $L$ = 10 mm. (b) At $L$ = 10\( \mu\mathrm{m} \). (c) At $L$ = 10 nm.}
  \label{fig:window_lif}
\end{figure}

\newpage
\clearpage

\textbf{(3) LiH:
}

\begin{figure}[htbp]
  \centering  \includegraphics[width=\linewidth,height=.65\textheight,keepaspectratio]{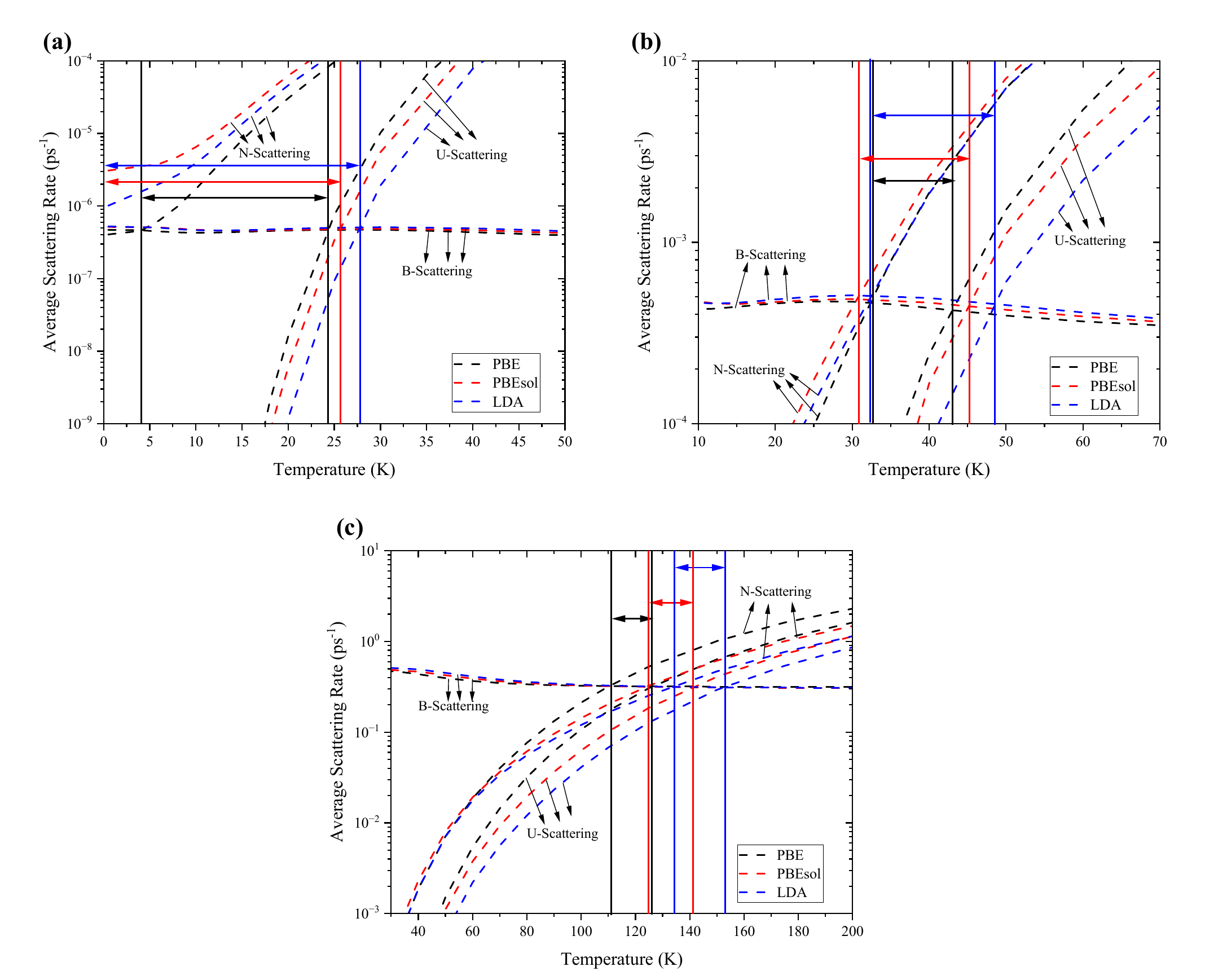}
  \caption{Phonon hydrodynamics windows for LiH. Calculations were done with PBE (black dashed lines), PBEsol (red dashed lines), and LDA (blue dashed lines). Double arrow lines show the window of phonon hydrodynamics for each functional. (a) At $L$ = 10 mm. (b) At $L$ = 10\( \mu\mathrm{m} \). (c) At $L$ = 10 nm.}
  \label{fig:window_lih}
\end{figure}

\newpage
\clearpage

\textbf{(4) NaH:
}

\begin{figure}[htbp]
  \centering  \includegraphics[width=\linewidth,height=.65\textheight,keepaspectratio]{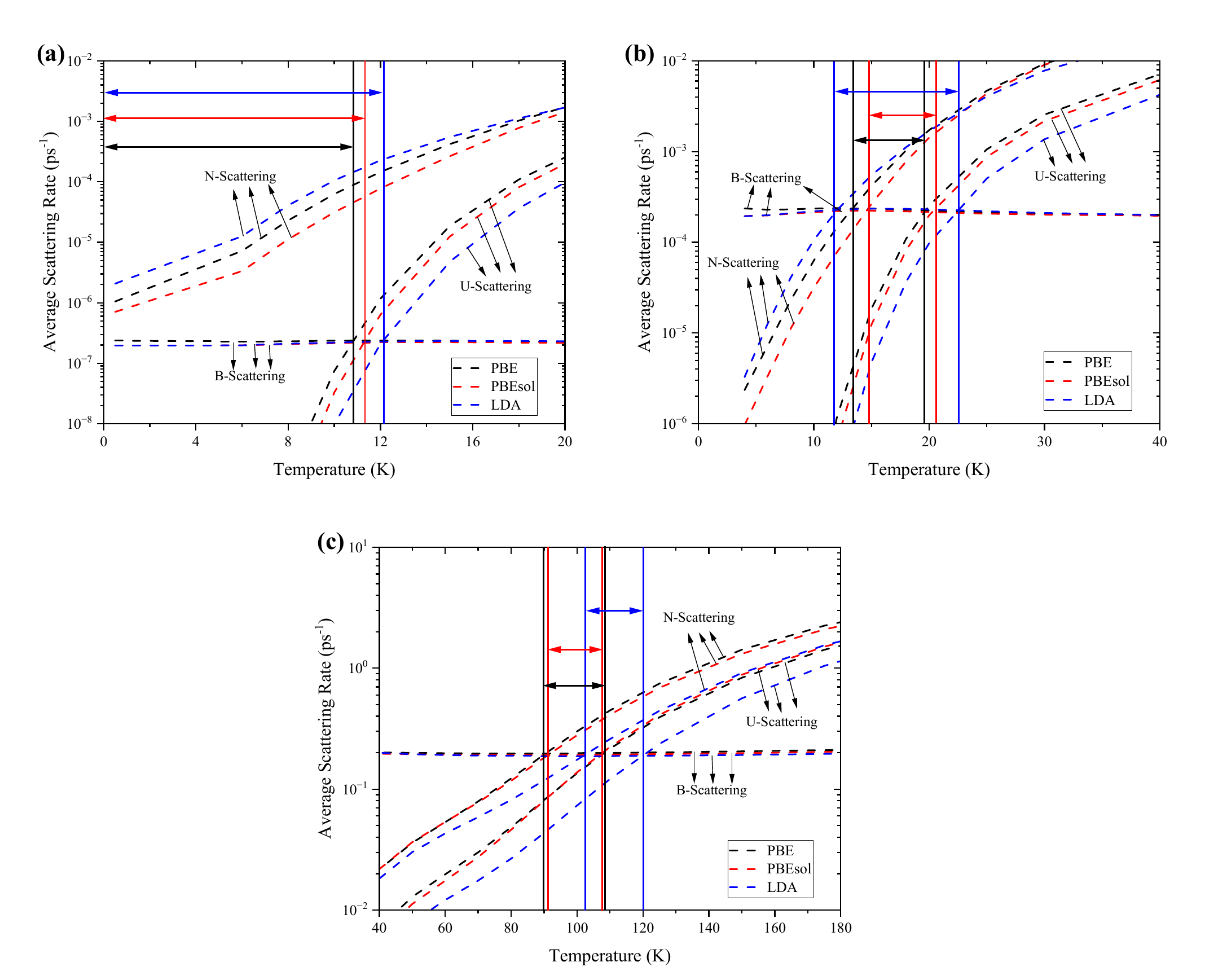}
  \caption{Phonon hydrodynamics windows for NaH. Calculations were done with PBE (black dashed lines), PBEsol (red dashed lines), and LDA (blue dashed lines). Double arrow lines show the window of phonon hydrodynamics for each functional. (a) At $L$ = 10 mm. (b) At $L$ = 10\( \mu\mathrm{m} \). (c) At $L$ = 10 nm.}
  \label{fig:window_nah}
\end{figure}


\newpage
\clearpage

\clearpage
\Needspace{0.90\textheight}
\section{Phonon hydrodynamics windows including isotope scattering}\label{sec:isotope_windows}

\begin{figure}[H]
  \centering
  \includegraphics[width=\linewidth,height=0.82\textheight,keepaspectratio]{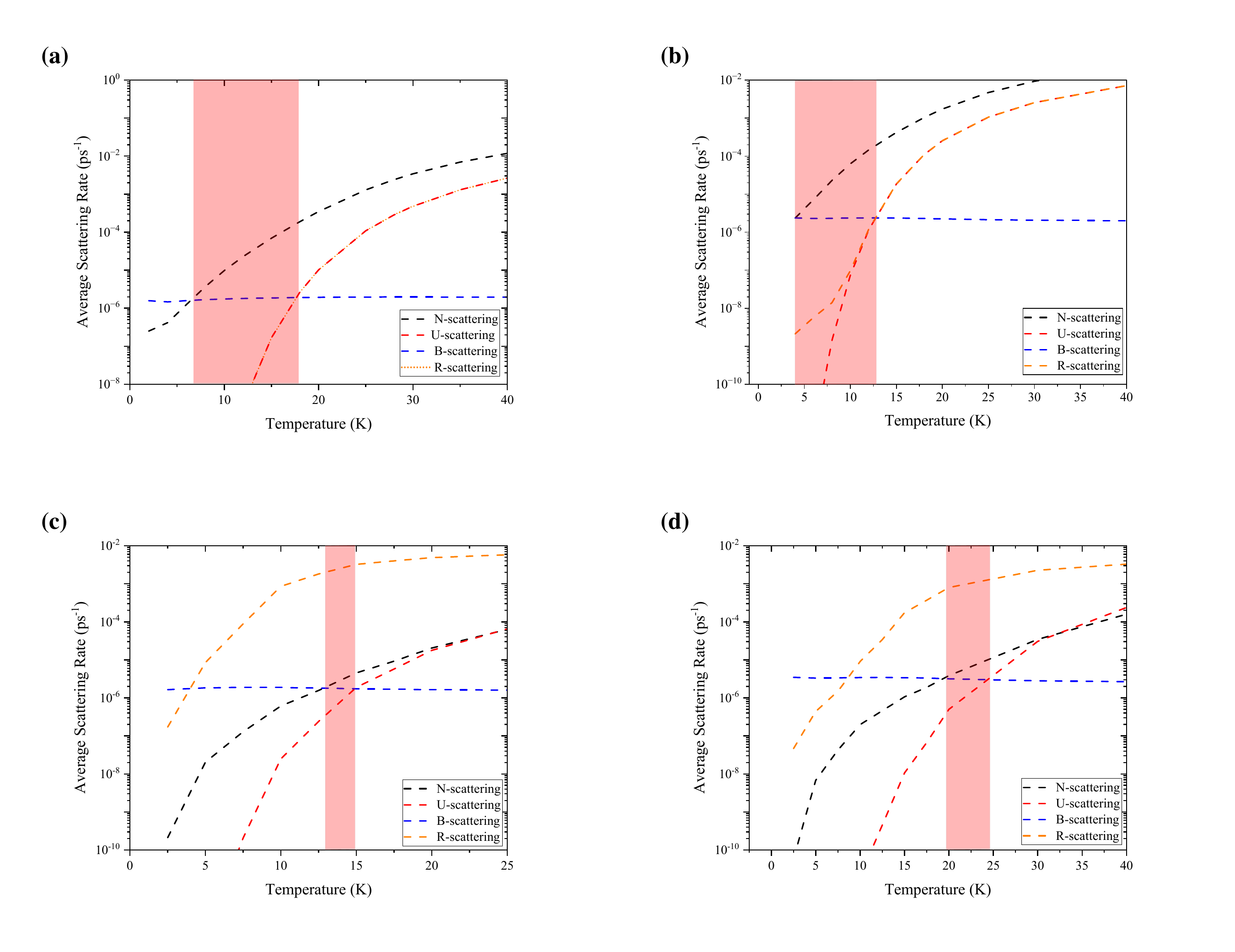}
  \caption{Phonon hydrodynamics windows for pure materials (shaded transparent red region) at $L$ = 1 mm for: (a) NaF using PBE. (b) NaH using PBE. (c) Ge (adapted from Ref.~\cite{huberman2022question, huberman2023revisiting}). (d) Si. Scattering rates are denoted by colors: N-scattering (black dashed lines), U-scattering (red dashed lines), B-scattering (blue dashed lines), and R-scattering (orange dashed lines).}
  \label{fig:isotope_windows}
\end{figure}


\newpage
\clearpage

\section{Thermal transport regimes including isotope scattering
}\label{sec:isotope_thermal}

\begin{figure}[htbp]
  \centering  \includegraphics[width=\linewidth,height=.65\textheight,keepaspectratio]{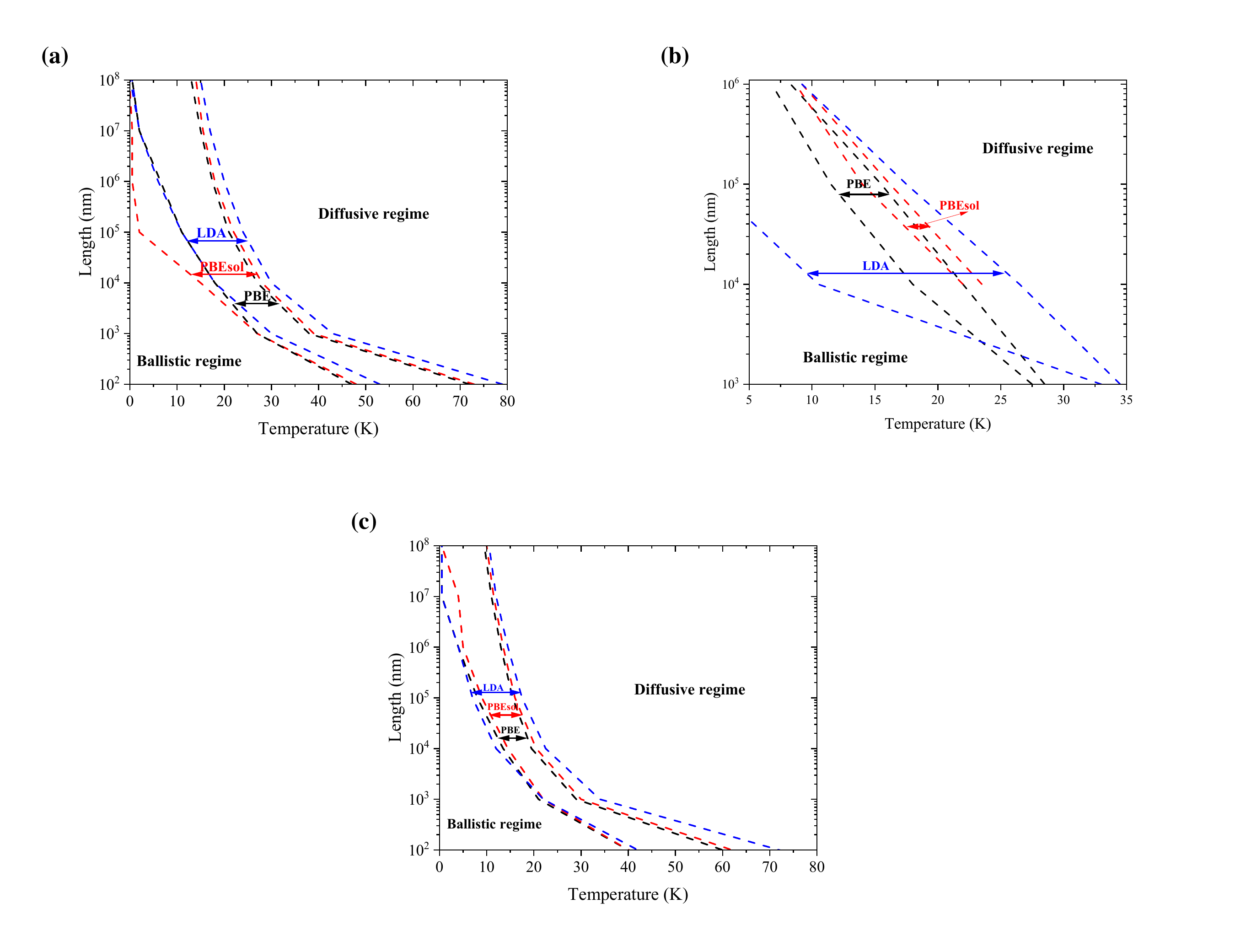}
  \caption{Thermal transport regimes: hydrodynamics regime, ballistic regime, and diffusive regime for (a) NaF, (b) LiF, and (c) NaH. Double arrow lines show the window of phonon hydrodynamics for each functional.}
  \label{fig:isotope_thermal}
\end{figure}


\newpage
\clearpage

\section{Direct solutions to the linearized BTE}\label{sec:LBTE}

\begin{figure}[htbp]
  \centering  \includegraphics[width=\linewidth,height=.35\textheight,keepaspectratio]{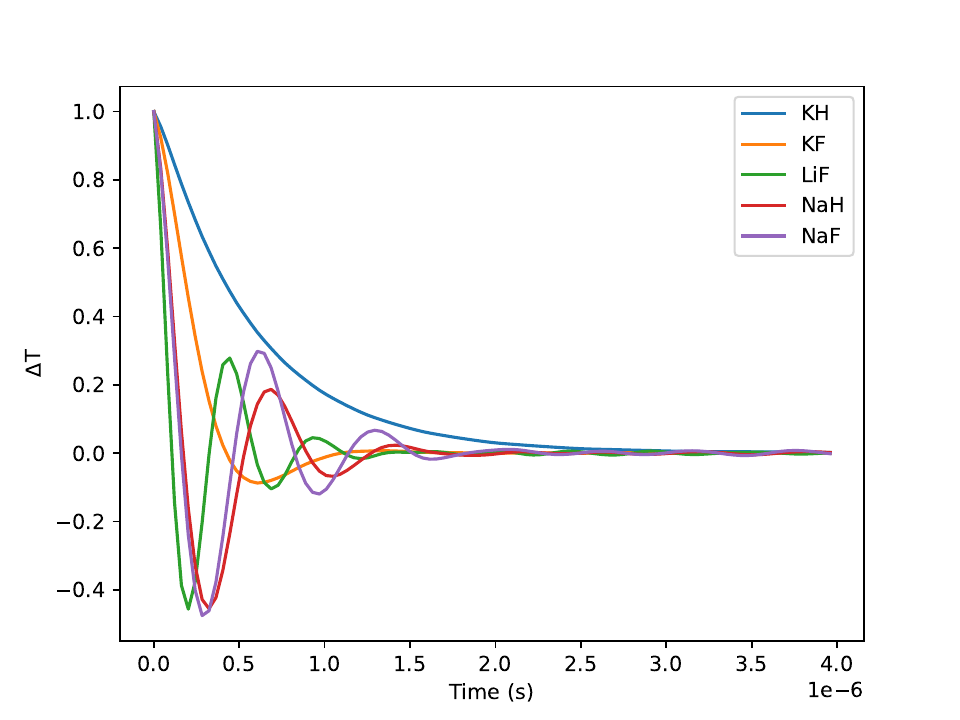}
  \caption{Temperature response for LiF, NaF, KF, KH, and NaH in the 1D-TTG for a grating period of 1.4 mm at $T$ = 15 K.}
  \label{fig:TTG}
\end{figure}

\begin{figure}[htbp]
  \centering  \includegraphics[width=\linewidth,height=.35\textheight,keepaspectratio]{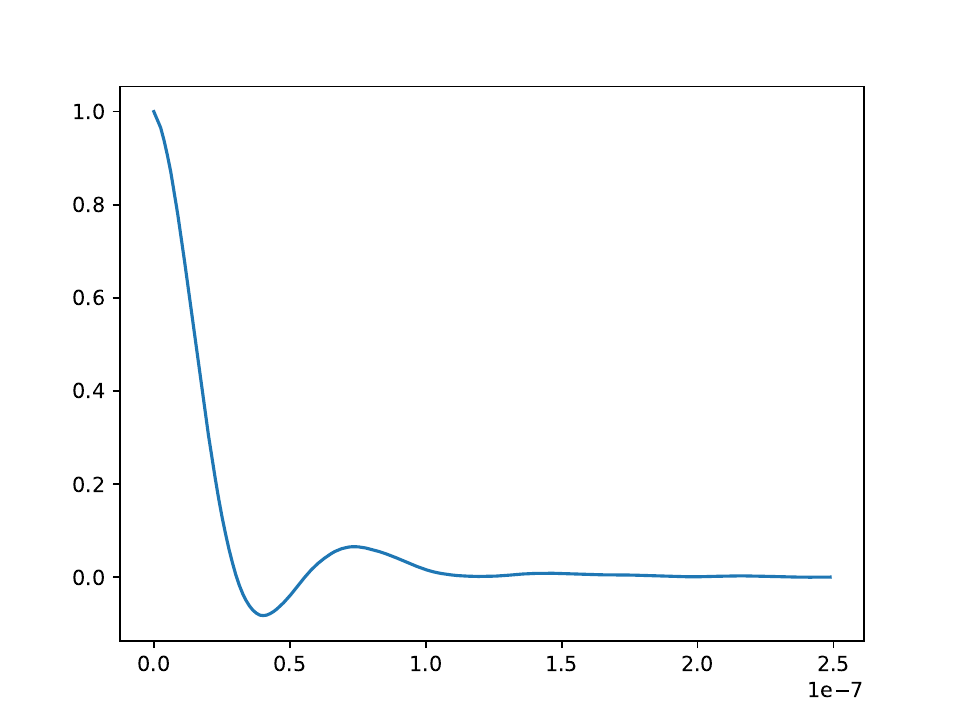}
  \caption{Temperature response for KH in the 1D-TTG for a grating period of 110 $\mu$m at $T$ = 15 K.}
  \label{fig:TTG2}
\end{figure}

We calculated the temperature response for the 1D transient thermal grating (TTG) geometry as described by Ref.~\cite{chiloyan2021green}. The oscillatory response in the temperature profile confirms the existence of second sound in LiF, NaF, KF, KH, and NaH, as shown in figures \ref{fig:TTG} and  \ref{fig:TTG2}. This result is consistent with experimental findings for NaF by Jackson \textit{et al.} \cite{jackson1970second}. To ensure consistency, the LDA functional was used for all materials.

\cleardoublepage
\bibliographystyle{iopart-num}

\bibliography{references}